\documentclass[preprint,showpacs,superscriptaddress]{revtex4}

\usepackage{graphicx}
\usepackage{dcolumn}
\usepackage{amsmath}
\usepackage[colorlinks=true, citecolor=blue, urlcolor = blue, linkcolor= red, bookmarks=true]{hyperref}

\begin{document}
%opening
\title{Particle acceleration of two general particles in the background of rotating Ay\'{o}n-Beato-Garc\'{i}a 
black holes}

\author{Fazlay Ahmed} 
\email{fazleyamuphysics@gmail.com}
\affiliation{Centre for Theoretical Physics,
 Jamia Millia Islamia,  New Delhi 110025
 India}
 
\author{Muhammed Amir}
\email{amirctp12@gmail.com}
\affiliation{Astrophysics and Cosmology Research Unit,
	School of Mathematics, Statistics and Computer Science,
	University of KwaZulu-Natal, Private Bag X54001,
	Durban 4000, South Africa}
 
\author{Sushant G. Ghosh}\email{sghosh2@jmi.ac.in}
\affiliation{Centre for Theoretical Physics, Jamia Millia Islamia,  New Delhi 110025 India}
\affiliation{Multidisciplinary Centre for Advanced Research and Studies (MCARS),\\ Jamia Millia Islamia, New Delhi 110025, India}
\affiliation{Astrophysics and Cosmology Research Unit,
	School of Mathematics, Statistics and Computer Science,
	University of KwaZulu-Natal, Private Bag X54001,
	Durban 4000, South Africa}

\begin{abstract}

The rotating Ay\'{o}n-Beato-Garc\'{i}a (ABG) black holes, apart from mass ($M$) and rotation parameter ($a$), has an additional charge  $Q$ and encompassed the Kerr black hole as particular case when $Q=0$. We demonstrate the ergoregions of rotating ABG black holes depend on both rotation parameter $a$ and charge $Q$, and the area of the ergoregions increases with increase in the values of $Q$, when compared with the Kerr black hole and the extremal regular black hole changes with the value of $Q$. Ban{\~a}dos, Silk and West (BSW) demonstrated that an extremal Kerr black hole can act as a particle accelerator with arbitrarily high center-of-mass energy ($E_{CM}$) when the collision takes place at any point in the ergoregion and thus in turn provides a suitable framework for Plank-scale physics. We study the collision of two general particles with different masses falling freely from rest in the equatorial plane of a rotating ABG black hole near the event horizon and find that the $E_{CM}$ of two colliding particles is arbitrarily high when one of the particles take a critical value of angular momentum in the extremal case, whereas for nonextremal case $E_{CM}$ for a pair of colliding particles is generically divergent at the inner horizon, and explicitly studying the effect of charge $Q$ on the $E_{CM}$ for ABG black hole. In particular, our results in the limit $Q\rightarrow 0$ reduce exactly to \emph{vis-$\grave{a}$-vis} those of the Kerr black hole. 
\end{abstract}


\maketitle

\section{Introduction}
\label{intro}
Ban\~{a}dos, Silk and West \cite{Banados:2009pr} have proposed that the collision of two particles, say dark 
matter particles, falling from rest at infinity into the Kerr black hole \cite{Kerr:1963ud} can produce an 
infinitely large center-of-mass energy ($E_{CM}$) when collision takes place in the vicinity of the event horizon, 
with the black hole maximally spinning, and one of the particle have critical angular momentum. This mechanism of 
particle acceleration by a black hole is particularly called BSW mechanism, which is interesting from the 
viewpoint of theoretical physics because new Planck scale physics is possible in the vicinity of the black holes. 
Further, the extremal Kerr black hole surrounded by dark matter could be regarded as a Planck scale collider, 
which might help us to explain the  astrophysical phenomena, such as the the active galactic nuclei and gamma ray 
bursts. Hence, the BSW mechanism about the collision of two particles near a rotating black hole has attracted 
significant attention \cite{Berti:2009bk,Jacobson:2009zg,Lake:2010bq,Wei:2010gq,Mao:2010di,Grib:2010dz,Liu:2010ja,Grib:2010xj,Zhu:2011ae, Hussain:2012zza,11,12,Igata:2012js,Zaslavskii:2012fh} (see also \cite{Harada:2011xz}, for a review). BSW phenomena is not only done for Kerr black holes but also for Kerr-Newman black holes 
\cite{2}, Regular Black holes \cite{Ghosh:2014mea,Pradhan:2014oaa,Amir:2015pja,Ghosh:2015pra}, higher dimensional 
black holes \cite{Tsukamoto:2013dna} and naked singularities \cite{4,5, 5a, Patil:2011ya,Patil:2011uf}. In 
particular for the Kerr-Newman black hole, the $E_{CM}$  of collision depends not only on the spin $ a $ but also 
on the charge $ Q $ of the black hole. Lake \cite{Lake:2010bq} analyzed the $E_{CM}$ of the collision occurring 
at the inner horizon of the non-extremal Kerr black hole and found that $E_{CM}$ is unlimited. Grib and Pavlov 
\cite{ Grib:2010xj} showed that the $E_{CM}$ for two particles collision can be unlimited even in the non-maximal 
rotation. Later, Zaslavskii \cite{Zaslavskii:2010jd,Zaslavskii:2010pw} demonstrated that an acceleration of 
the particles by BSW method is a universal property of rotating black holes. Recently, Harada and Kimura 
\cite{Harada:2011xz} generalized the BSW analysis of two colliding particles to general geodesics in the Kerr 
black hole to show an arbitrarily high  $E_{CM}$ can occur near the horizon of maximally rotating black holes. 
Further, the subject of particle acceleration for two different masses and energetic particles is extended for a 
class of black holes \cite{Liu:2011wv,Amir:2016nti}. The analysis is also valid to the collision of particles in 
non-equatorial motion for Kerr black holes \cite{6} and Kerr-Newman black holes \cite{Liu:2011wv}. The general 
explanation of BSW phenomenon is proposed in \cite{9}. 

It turns out that the horizon structure of the rotating regular black hole is complicated as compared to the Kerr 
black hole, which depend on the mass and spin of the black hole and on an additional deviation parameter that 
measures potential deviations from the Kerr metric, and includes the Kerr metric as the special case if this 
deviation parameter vanishes. Further, these regular black holes are very important as astrophysical black holes, 
like Cygnus X-1 or Sgr$  A^* $, although suppose to be like the Kerr black hole \cite{32, 33}, but the definite 
nature of astrophysical black hole still need to be tested, and it may deviate from the standard Kerr 
black hole. More recently, the BSW mechanism when applied to an extremal regular black holes 
\cite{Ghosh:2014mea, Pradhan:2014oaa}, also lead to divergence of the $E_{CM}$. Hence, the BSW mechanism should 
be suitably adapted for the rotating regular black hole, which has very complicated horizon structure as 
compared to the Kerr black hole and can have extremal black holes depending on the additional parameter 
\cite{Ghosh:2014mea,Amir:2015pja,Ghosh:2015pra}. 
 
The rotating regular Ay\'{o}n-Beato-Garc\'{i}a (ABG) black hole \cite{Bambi:2013ufa} is an exact solution of 
Einstein's equations coupled to nonlinear electrodynamics. The rotating ABG black holes are axisymmetric, 
asymptotically flat, and depend on the mass ($M$) and spin ($a$) of the black hole as well as on a charge ($ Q $) 
that measures potential deviations from the Kerr metric and includes the Kerr metric as the special case if this 
deviation parameter vanishes.
 
The main goal of this paper is to  discuss the detailed  behavior of the $E_{CM}$ for two particles with different 
rest masses $ m_1 $ and $ m_2 $, falling freely from rest at infinity in the background of a rotating ABG black 
hole and calculate the $E_{CM}$. We go on to show that the $E_{CM}$ near the horizon of an extremal ABG black hole 
is arbitrarily high when one of the two particles acquire the critical angular momentum and also high at the inner 
horizon. We also explicitly show the effect of the parameter $ Q $ on BSW mechanism and ergoregion of the black 
hole. 

The paper is organized as follows. In Sec.~\ref{sptm}, we shall discuss horizons of the rotating ABG spacetime and 
also demonstrate the effect of parameter $Q$ on ergoregion. In Sec.~\ref{geqm}, we will discuss the equations of  
motion for a particle in the background of a rotating ABG black hole. The calculation of expression for $E_{CM}$ 
of the collision for two general particles and  their  properties is the subject of  Sec.~\ref{cme} and  we 
conclude in the Sec.~\ref{con}. 

\section{Rotating Ay\'{o}n-Beato-Garc\'{i}a Black Holes} \label{sptm}
The spherically symmetric ABG black hole is an exact regular solution of general relativity with nonlinear electrodynamics field as a source, and it satisfies the weak energy condition. The gravitational field of ABG solution is described by the metric \cite{AyonBeato:1998ub}:
\begin{eqnarray}\label{abg}
ds^2 &=& -f(r)dt^2+\frac{1}{f(r)}d r^2+ r^2 d \Omega^2,
\end{eqnarray}
with
\begin{eqnarray}
f(r)=1-\frac{2 M r^2}{(r^2+Q^2)^{3/2}}+\frac{Q^2 r^2}{(r^2+Q^2)^{2}},
\end{eqnarray}
and $d \Omega^2=d \theta^2+\sin^2 \theta d \phi^2$. The parameter $M$ and $Q$ are respectively, mass and electric 
charge. The associated electric field is 
\begin{eqnarray}
E=Q r^4\left[\frac{r^2-5Q^2}{(r^2+Q^2)^4}+\frac{15}{2}\frac{M}{(r^2+Q^2)^{7/2}} \right].
\end{eqnarray}
Note that the ABG black hole is asymptotically go over to Reissner-Nordst{\"o}rm black hole
\begin{eqnarray}
f(r) &=& 1-\frac{2M}{r}+\frac{Q^2}{r^2}+O(1/r^3), \nonumber \\
E &=& \frac{Q}{r^2}+O(1/r^3).\nonumber
\end{eqnarray}
This solution corresponds to a regular charged black hole when $|Q|\leq 0.6M$, with the curvature invariant and electric field regular everywhere including at $r=0$ \cite{AyonBeato:1998ub}.

The rotating ABG black hole metric obtained in \cite{Toshmatov:2014nya}, beginning with the metric (\ref{abg}) 
and applying the Newman-Janis transformation, they constructed a rotating ABG metric. The gravitational field of rotating ABG spacetime is 
described by a metric which in the Boyer-Lindquist
coordinates  (with $G=c=1$) \cite{Toshmatov:2014nya} reads:
\begin{eqnarray}\label{metric}
ds^{2} &=& -f(r,\theta)dt^{2} - 2 a \sin^{2}\theta (1-f(r,\theta))d\phi dt +\frac{\Sigma}{\Delta}dr^2  \nonumber \\ 
& + & \Sigma d \theta^{2} + \sin^{2}\theta [\Sigma-a^{2}(f (r,\theta)-2)\sin^{2}\theta]d \phi^{2},
\end{eqnarray}
where, $\Sigma = r^2 + a^{2}\cos^{2}\theta, \;\;\;\; \Delta=\Sigma f(r,\theta) +a^{2}\sin^{2}\theta$, \; \;
$a$ is a rotation parameter and the function $f(r, \theta)$ is given by
\begin{eqnarray}\label{f}
f(r,\theta)=1-\frac{2 M r \sqrt{\Sigma} }{(\Sigma +Q^2)^{3/2}}+\frac{Q^2 \Sigma}{(\Sigma +Q^2)^2}.
\end{eqnarray}

The ABG metric~(\ref{metric}) is a regular rotating charged black hole \cite{Ghosh:2014mea,Toshmatov:2014nya}, which go over to Kerr black holes  \cite{Kerr:1963ud} when $Q=0$. When $a=0$, it reduces to the ABG black hole \cite{AyonBeato:1998ub}, and for $a= Q =0$, it reduces to the Schwarzschild black hole \cite{schw}. The invariant Ricci scalar ($R_{ab} R^{ab}$ ) and Kretschmann scalar ($R_{abcd} R^{abcd}$ ) are suppose to be regular everywhere including at $r=0$ \cite{Ghosh:2014mea}. The metric~(\ref{metric}) becomes singular if $\Sigma=0$ or $\Delta=0$ whereas $\Sigma=0$ is the curvature singularity and $\Delta=0$ is the coordinate singularity. $\Delta=0$ gives horizons of the rotating ABG black holes \cite{Ghosh:2014mea}.

\begin{figure*}
\begin{tabular}{c c c c}
\includegraphics[width=0.245\linewidth]{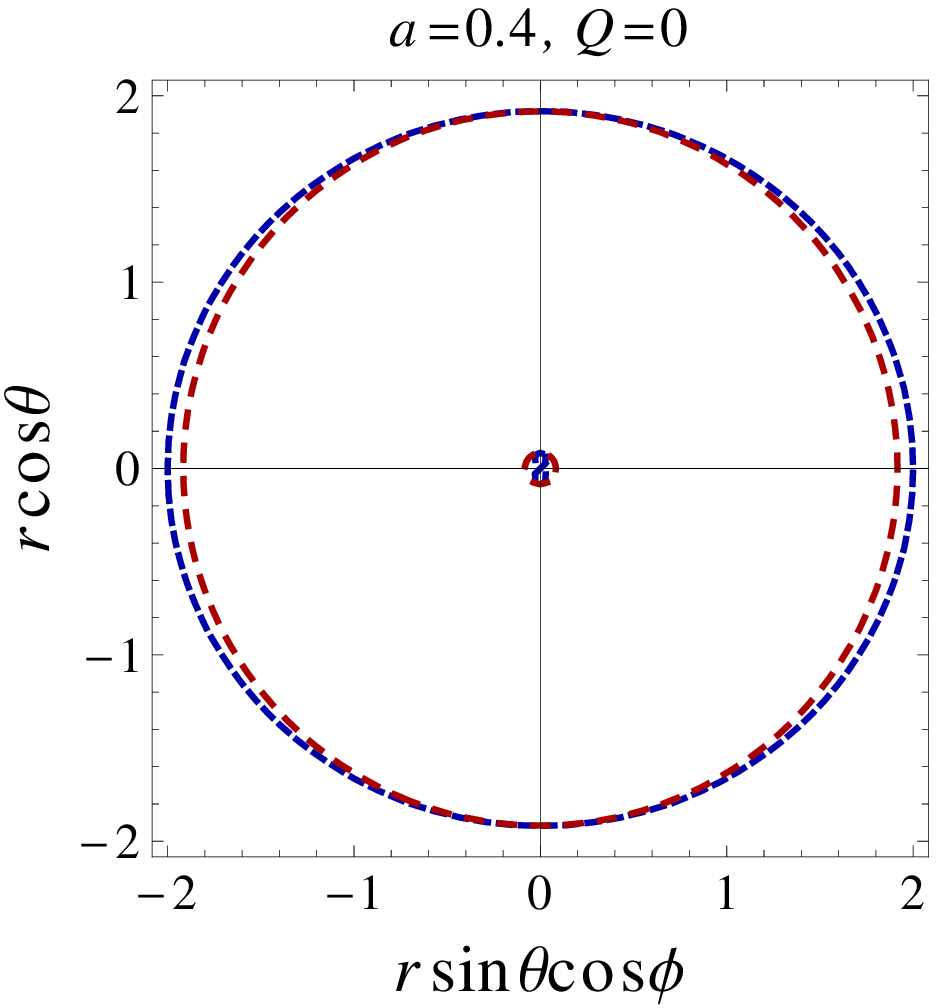}
\includegraphics[width=0.245\linewidth]{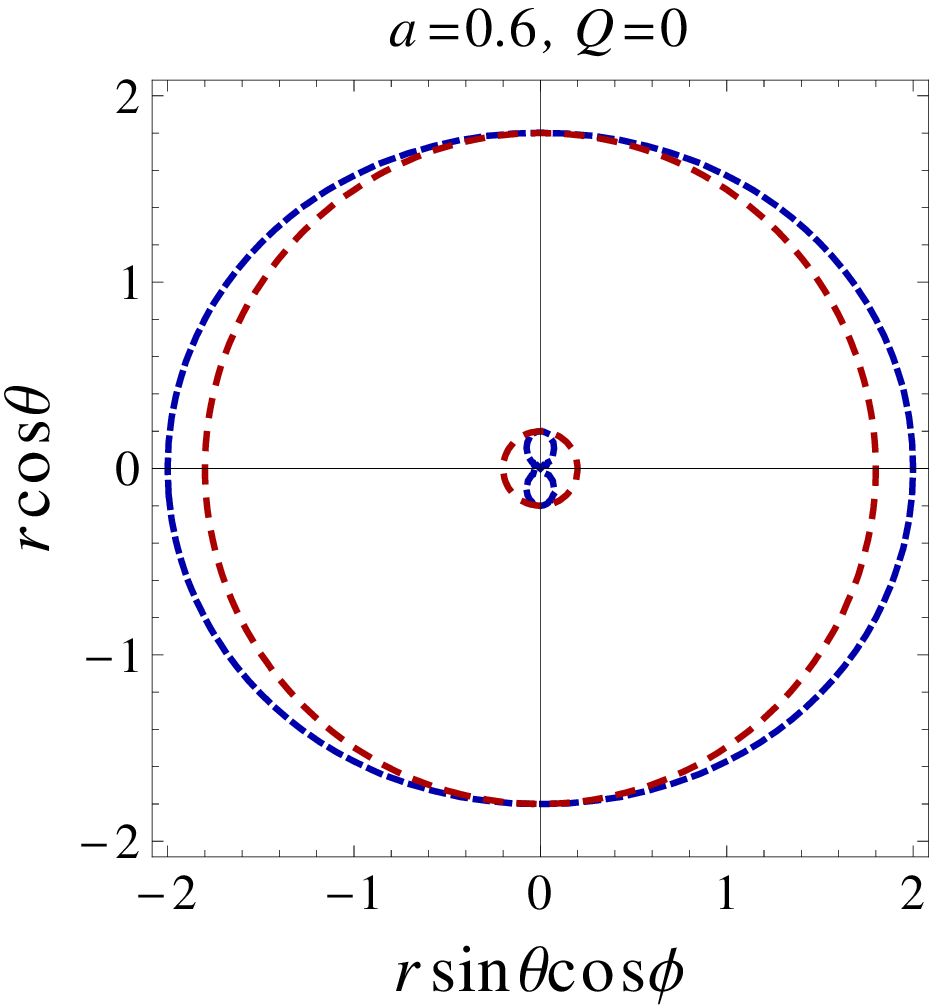}
\includegraphics[width=0.245\linewidth]{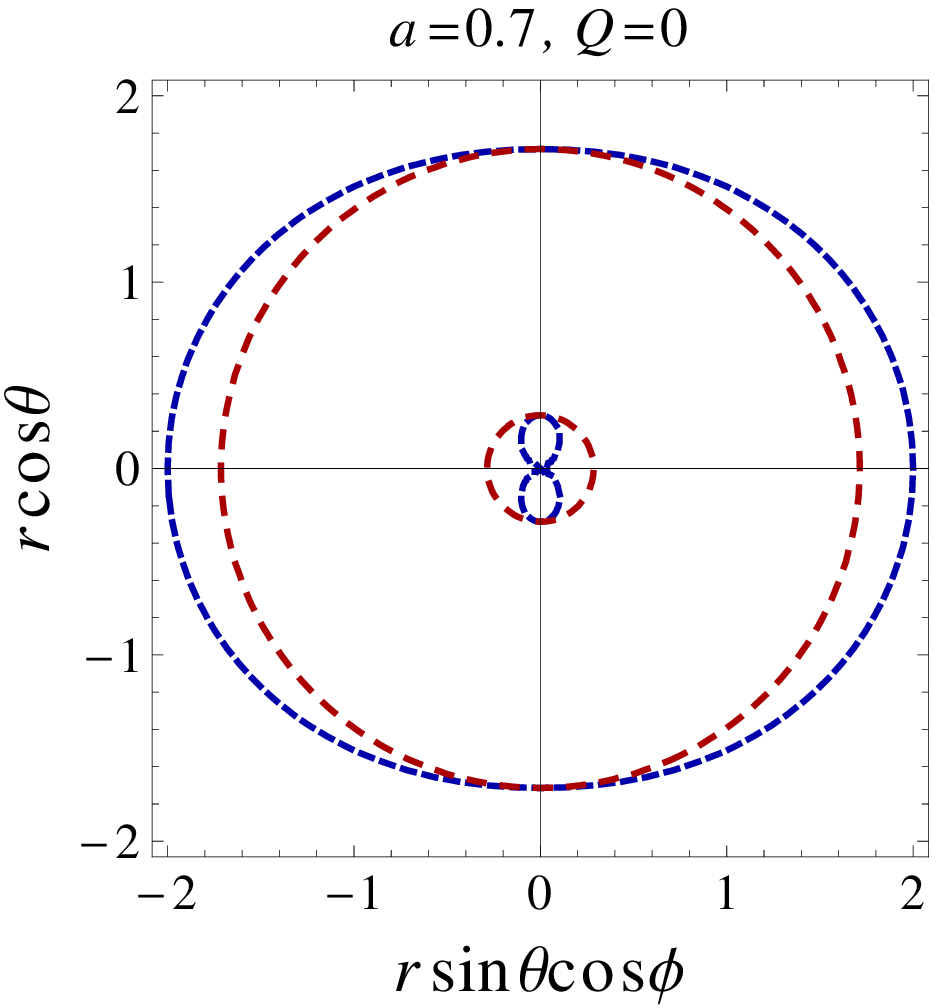}
\includegraphics[width=0.245\linewidth]{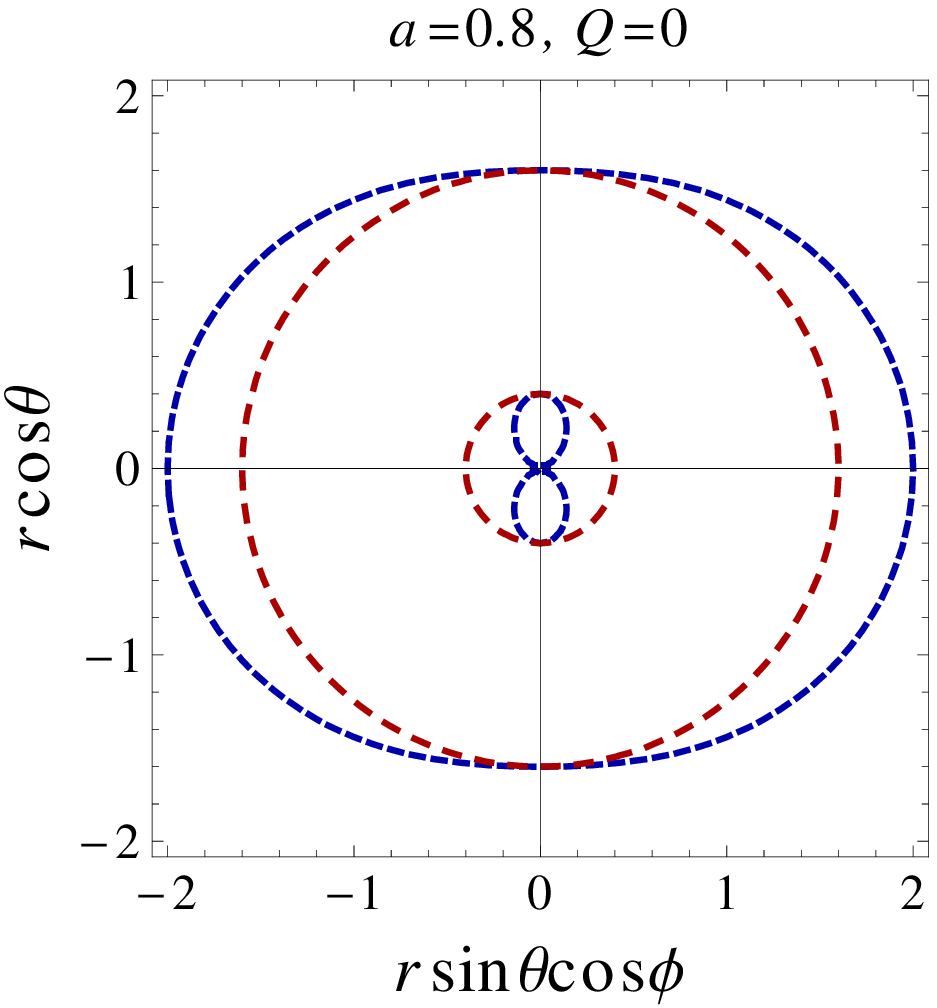}
\end{tabular}
\caption{\label{figA} Plots showing the behavior of ergoregion in the xz-plane of rotating ABG black hole for  $Q=0$ and different values of $a$. The blue and red lines correspond to the static limit surface and horizons.}
\end{figure*}

\subsection{Horizons and ergoregions of rotating ABG black holes}
In this section, we would like to study the properties of an ergoregion of a rotating ABG black hole which is the 
region between the static limit surface and the event horizon. In fact, the ergoregion plays an important role in 
the astrophysics for a possible observational effects, since in this region the Hawking radiation can be analyzed.  
The ergoregion is also important due to energetics of black holes and also Penrose process \cite{Penrose:1971uk}.  
The static limit surface ($r_{+}^{sls}$) is also known as infinite redshift surface, where the time-translation 
Killing vector becomes null. Another property of the static limit surface is that the time-like geodesics becomes 
space-like  after crossing the static limit surface. The static limit surface of a rotating ABG black hole can be 
calculated by solving the equation $g_{tt}= f(r,\theta)=0$.

In the case of $Q=0$, it represents the static limit surface of a Kerr black hole. The metric has a coordinate 
singularity at $\Delta=0$. Horizons of the metric (\ref{metric}) are given by $g^{rr}=\Delta=0$, i.e., 
\begin{equation}\label{eh}
 \Sigma f(r, \theta) + a^2 \sin^{2} \theta  = 0.
 \end{equation}
An ergoregion is located outside an event horizon, which has an oblate shape and touches the event horizon at 
poles and it has a highest radius at the equator. It turns out that Eq.~(\ref{eh}) admits two positive roots which 
take different values with charge $Q$. The two roots corresponding to the horizons of the black hole. Let us 
assume that $r^{EH}_{+}$ and $r^{CH}_{-}$ define respectively, the outer horizon (event horizon) and inner horizon 
of the black hole, and when they are equal  $r^{EH}_{+}= r^{CH}_{-} = r_{ex}$, corresponds to an extremal black 
holes with degenerate horizons. Inside the ergoregion an object can not remain static, but it moves in the 
direction of black hole spin. In ergoregion, an object can enter and escape, and we can extract energy and mass 
via the Penrose process \cite{Penrose:1971uk}.

\begin{widetext}
\begin{table}
\caption{Table for different values of $a$ and $Q$ for ABG black hole. Parameter $\delta^{a}$ is the region between static limit surface and event horizon ($\delta^{a}=r_{+}^{sls}-r^{EH}_{+}$).}
\begin{center}\label{Table A}
\begin{tabular}{| c| c c c | c c c | c c c |c c c|}
\hline
 &\multicolumn{3}{c|}{$a=0.5$}  &  \multicolumn{3}{c|}{$a=0.6$} &  \multicolumn{3}{c|}{$a=0.7$} &  \multicolumn{3}{c|}{$a=0.8$}\\
\hline
$ Q $ & $r^{EH}_{+}$ & $r_{+}^{sls}$  &$\delta^{a}$ & $r^{EH}_{+}$ & $r_{+}^{sls}$  & $\delta^{a}$
& $r^{EH}_{+}$ & $r_{+}^{sls}$  & $\delta^{a}$ & $r^{EH}_{+}$ & $r_{+}^{sls}$  &$\delta^{a}$ \\
\hline  
0.0  &  1.86603  &  1.93541  &  0.06938  &  1.80000  & 1.90554  & 0.10554 & 1.71414 & 1.86891  &  0.15477 & 1.60000 & 1.82462 &  0.22462 \\
0.1  &  1.85116  &  1.92196  &  0.07080  &  1.78371  & 1.89163  & 0.10792 & 1.69558 & 1.85440  &  0.15882 & 1.57732 & 1.80932 & 0.23200 \\
0.2  &  1.80488  &  1.88037  &  0.07549  &  1.73257  & 1.84856  & 0.11599 & 1.63641 & 1.80935  &  0.17294 & 1.50259 & 1.76161 & 0.25902 \\ 
0.3  &  1.72097  &  1.80640  &  0.08543  &  1.63779  & 1.77155  & 0.13376 & 1.52186 & 1.72821  &  0.20635 & 1.33877 & 1.67476 & 0.33599 \\
0.4  &  1.58263  &  1.68988  &  0.10725  &  1.47161  & 1.64881  & 0.17720 & 1.28210 & 1.59665  &  0.31455 &    -    & 1.53012 &    -    \\
\hline   
\end{tabular}
\end{center}
\end{table}
\end{widetext}

\begin{figure*}
   \begin{tabular}{c c c c}
\includegraphics[width=0.245\linewidth]{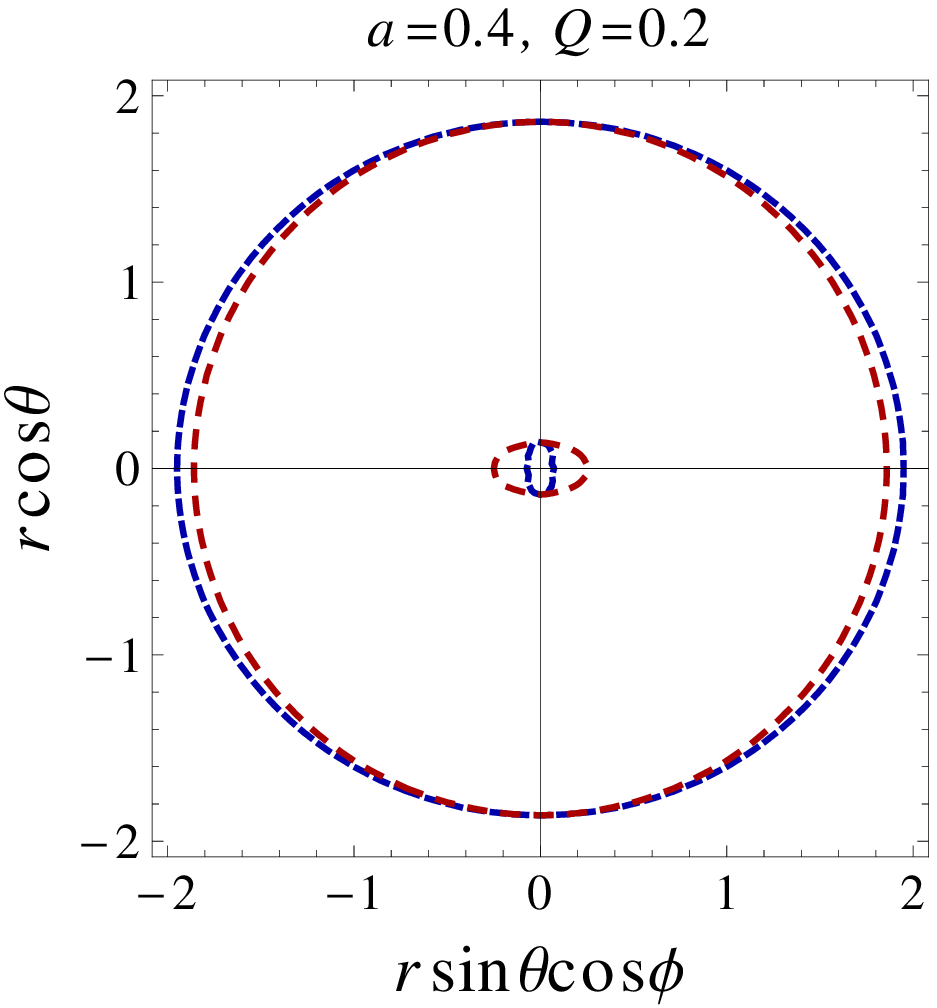}
\includegraphics[width=0.245\linewidth]{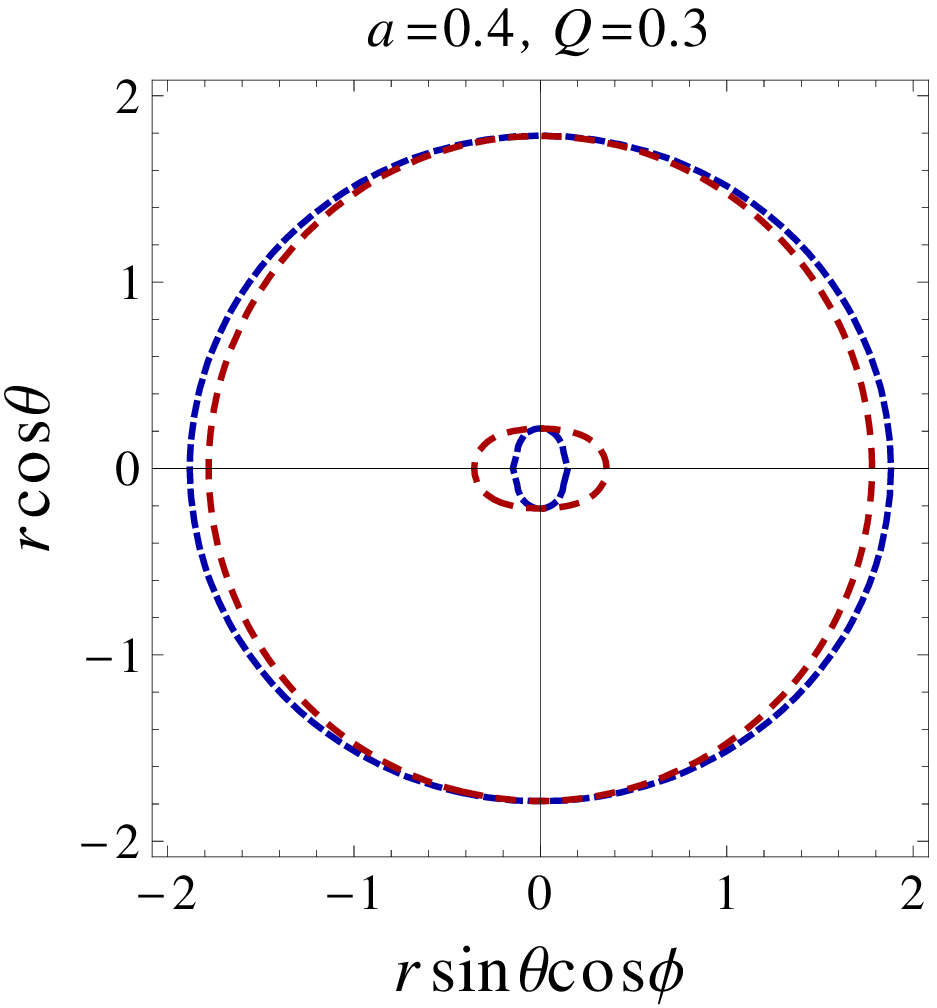}
\includegraphics[width=0.245\linewidth]{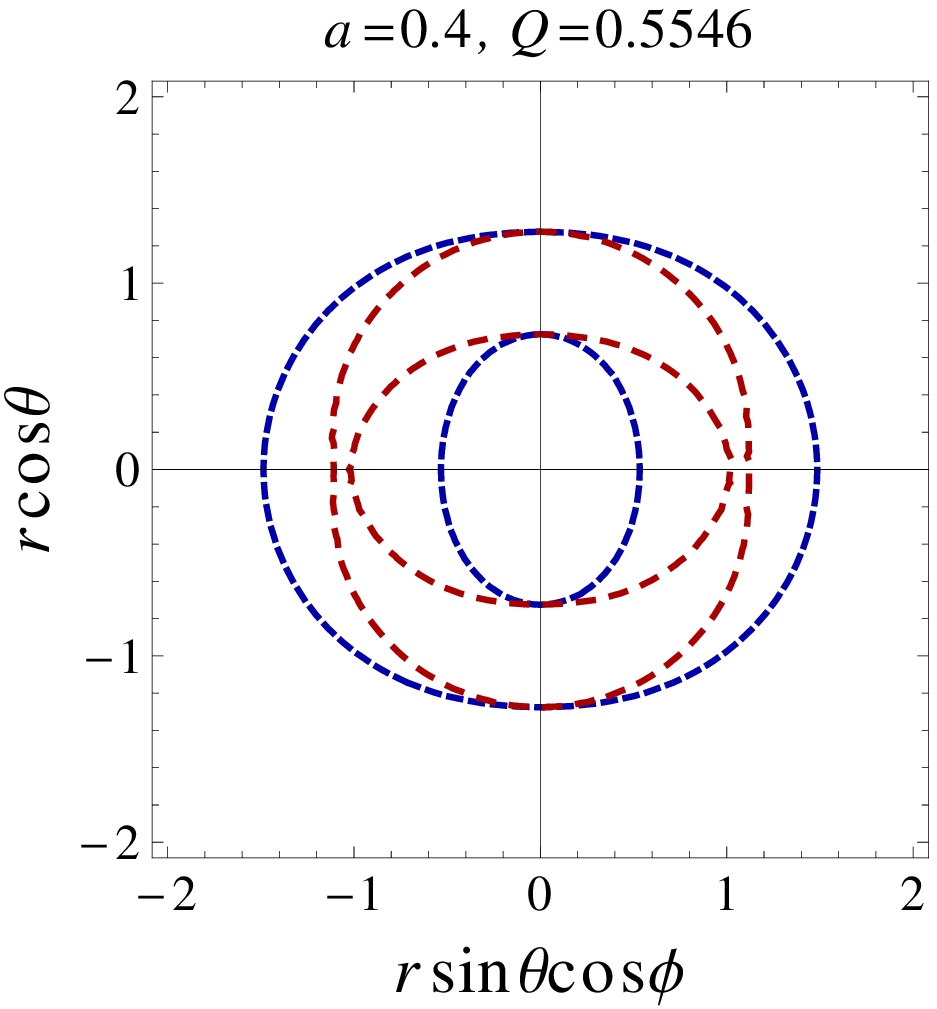}
\includegraphics[width=0.245\linewidth]{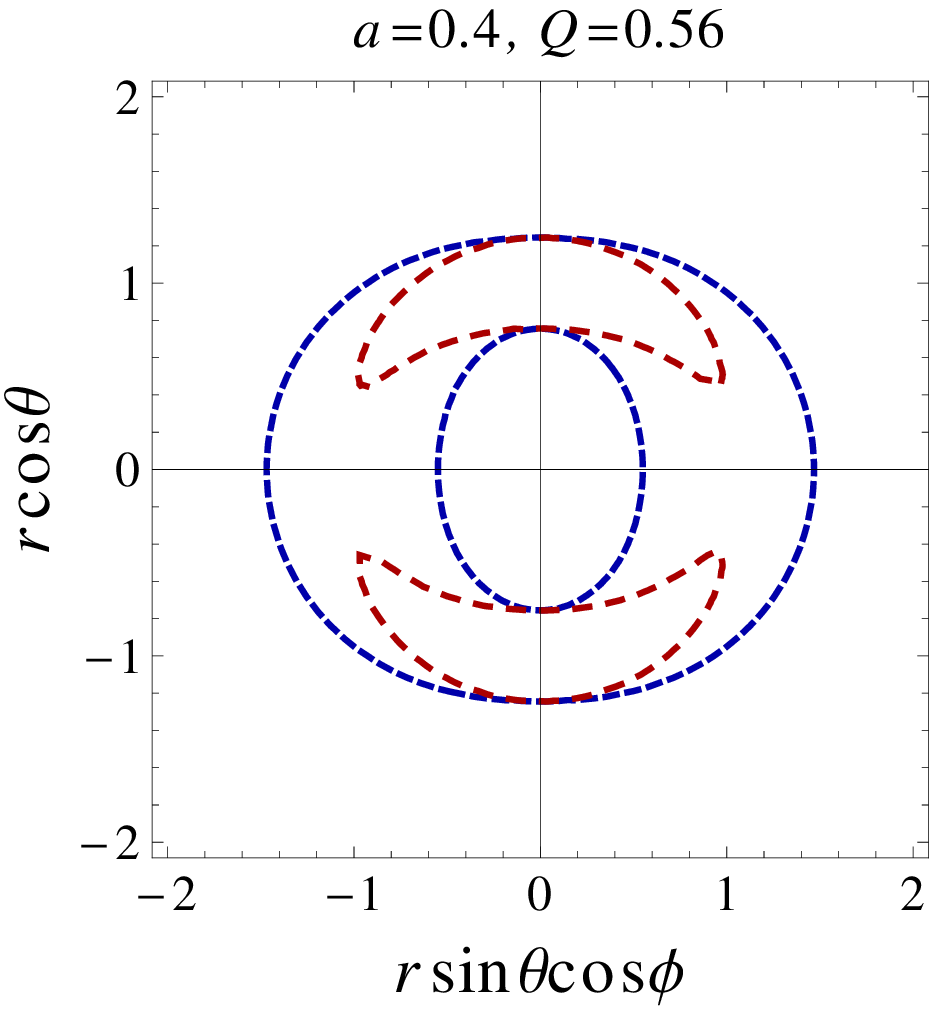}\\
\includegraphics[width=0.230\linewidth]{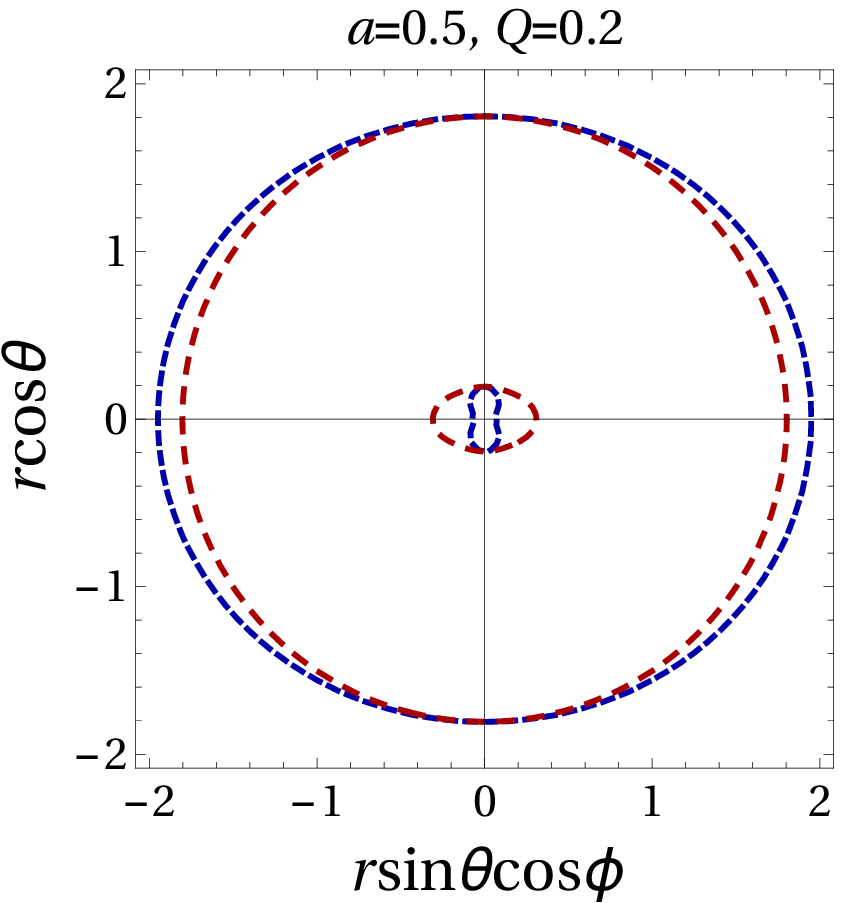}
\includegraphics[width=0.245\linewidth]{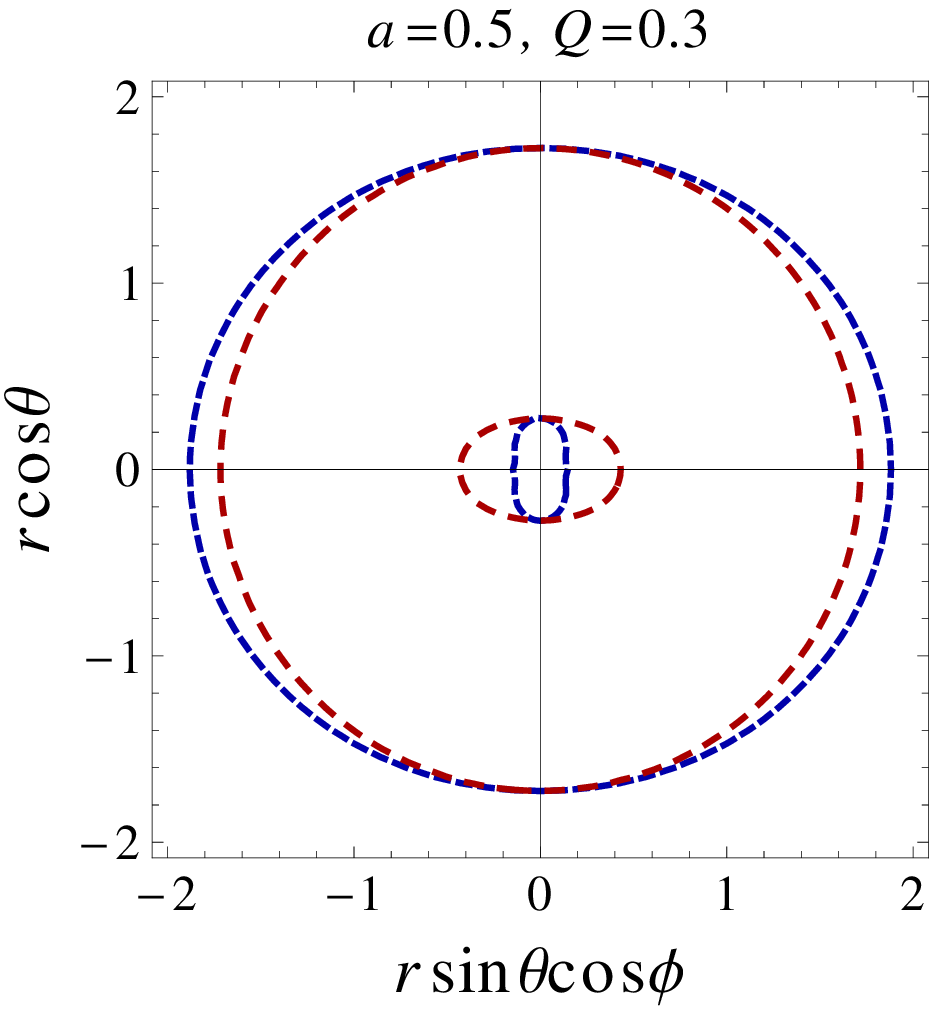}
\includegraphics[width=0.245\linewidth]{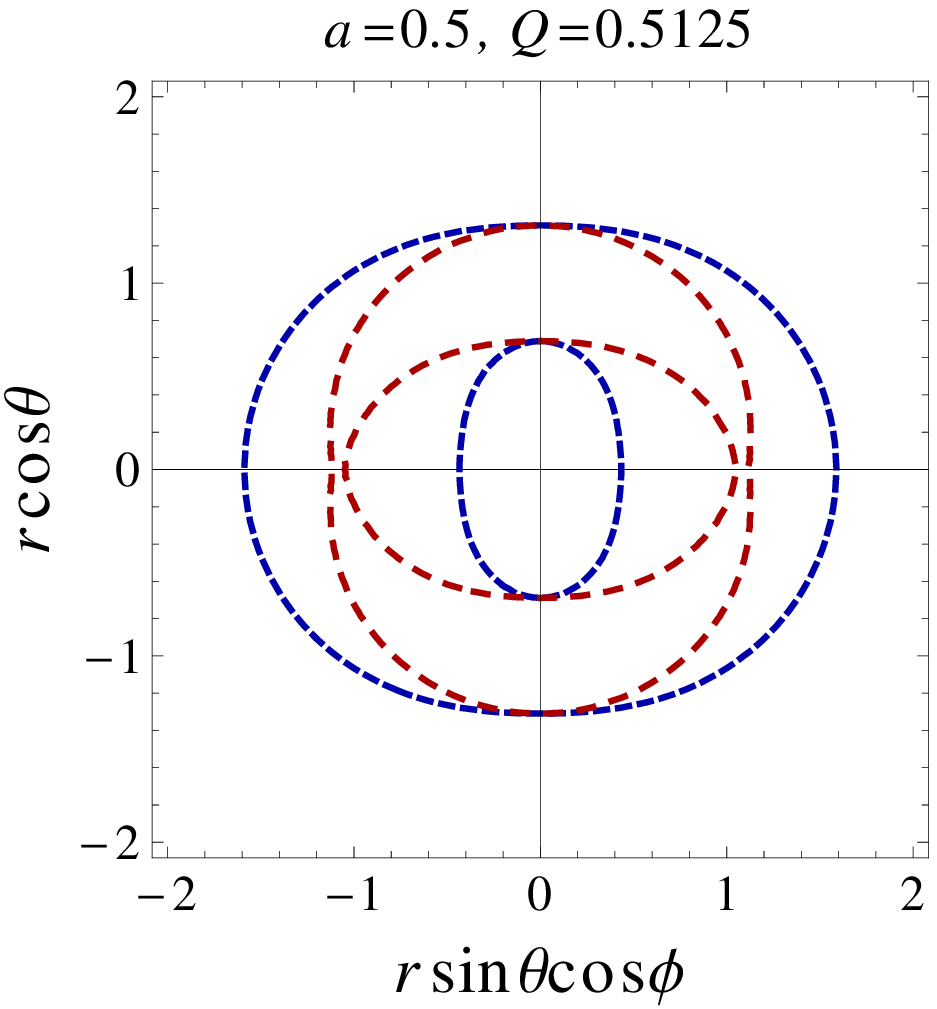}
\includegraphics[width=0.245\linewidth]{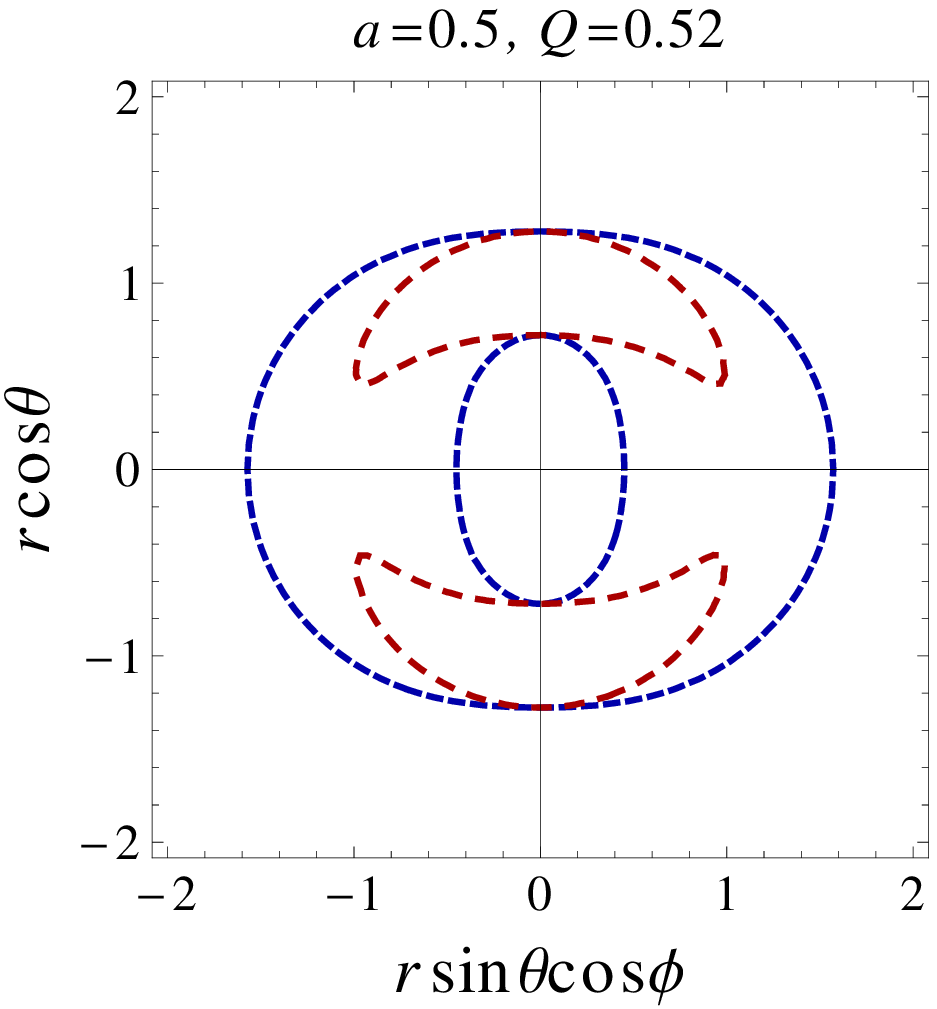}\\
\includegraphics[width=0.245\linewidth]{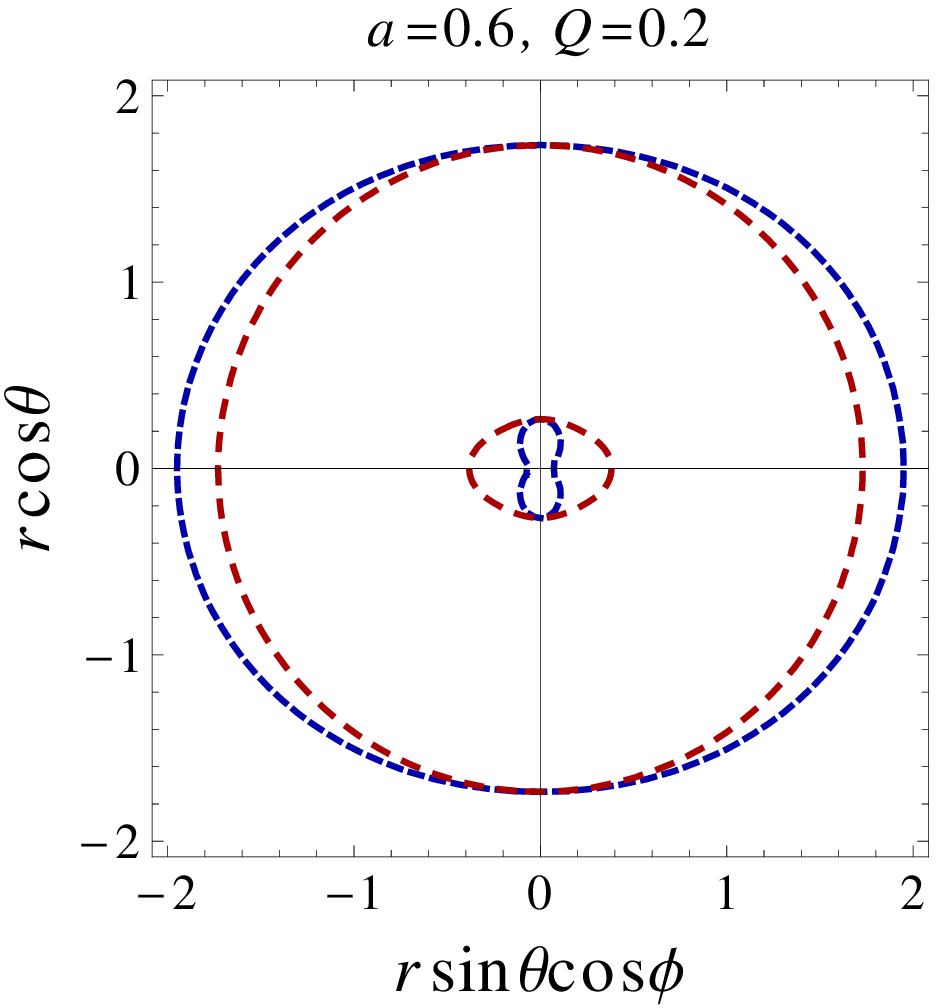}
\includegraphics[width=0.245\linewidth]{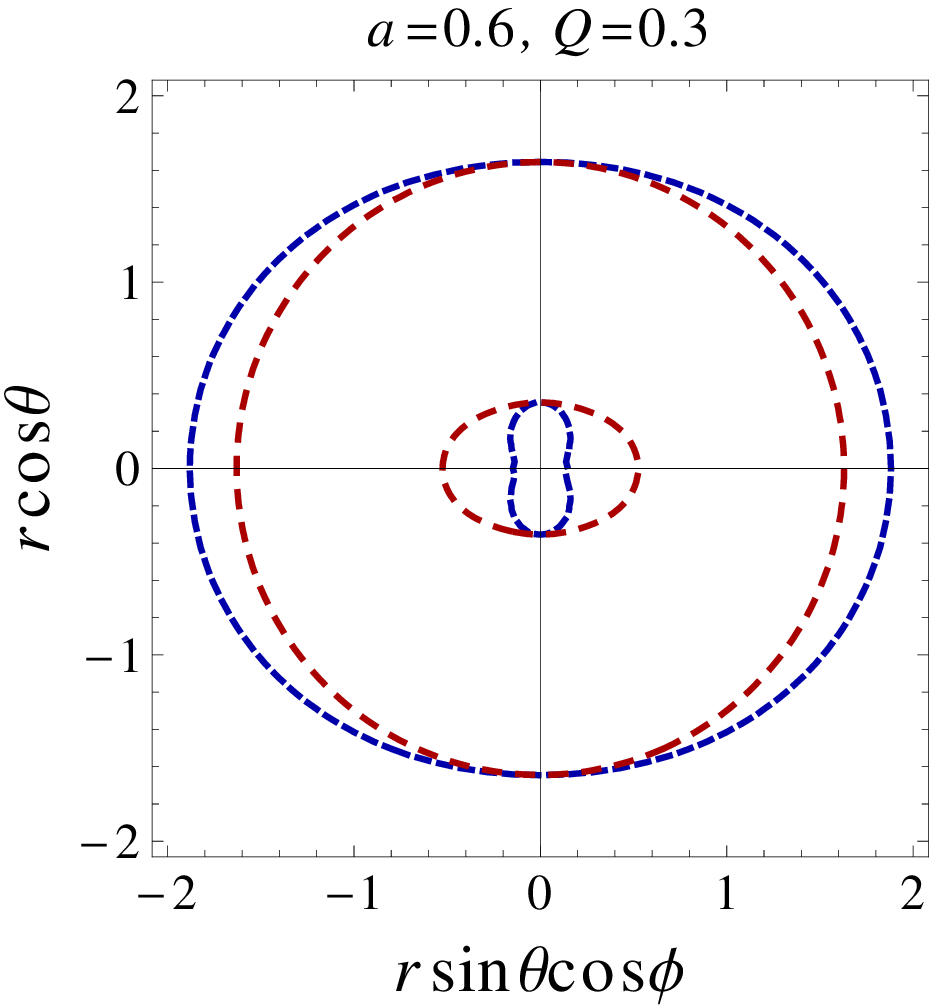}
\includegraphics[width=0.245\linewidth]{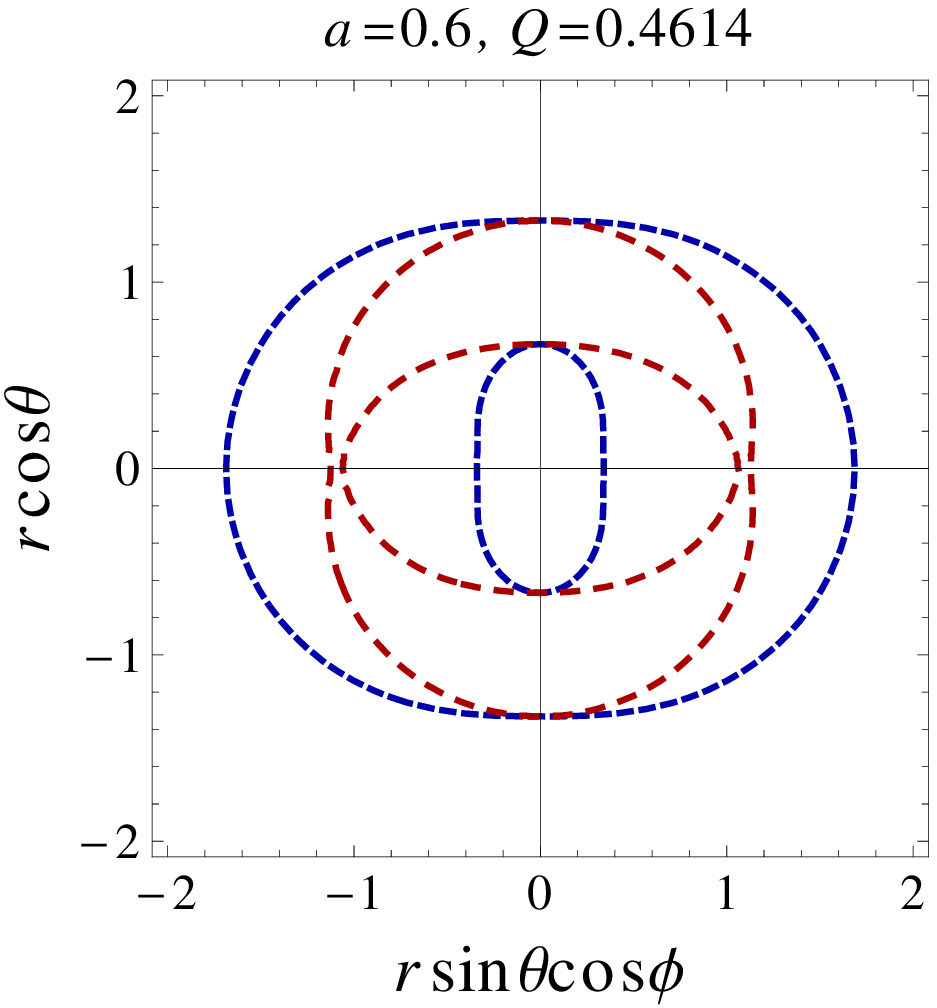}
\includegraphics[width=0.245\linewidth]{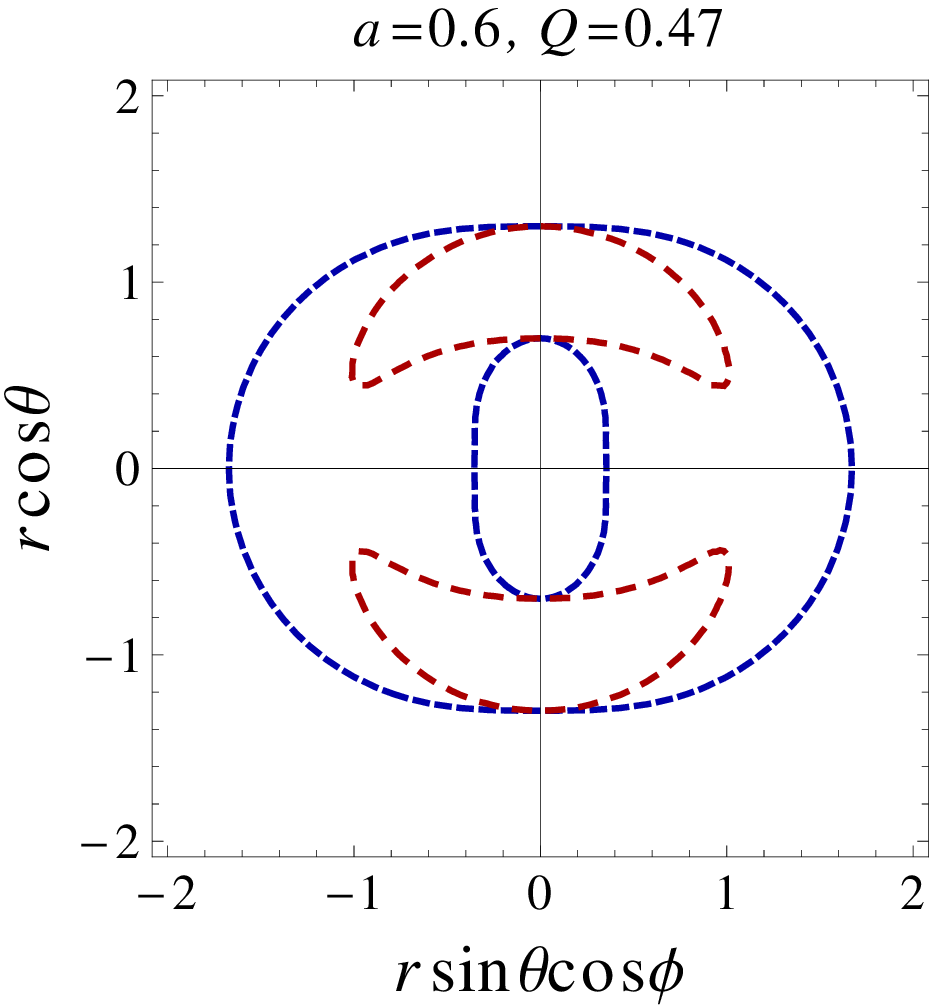}\\
\includegraphics[width=0.245\linewidth]{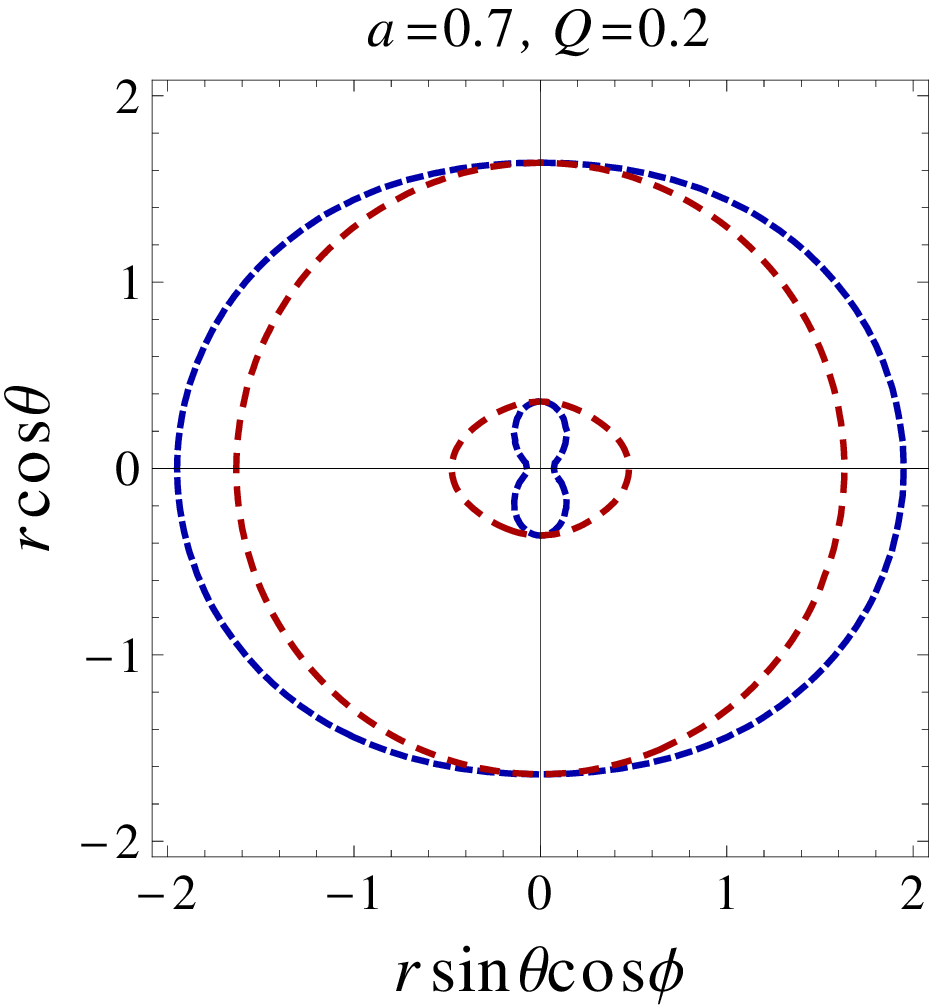}
\includegraphics[width=0.245\linewidth]{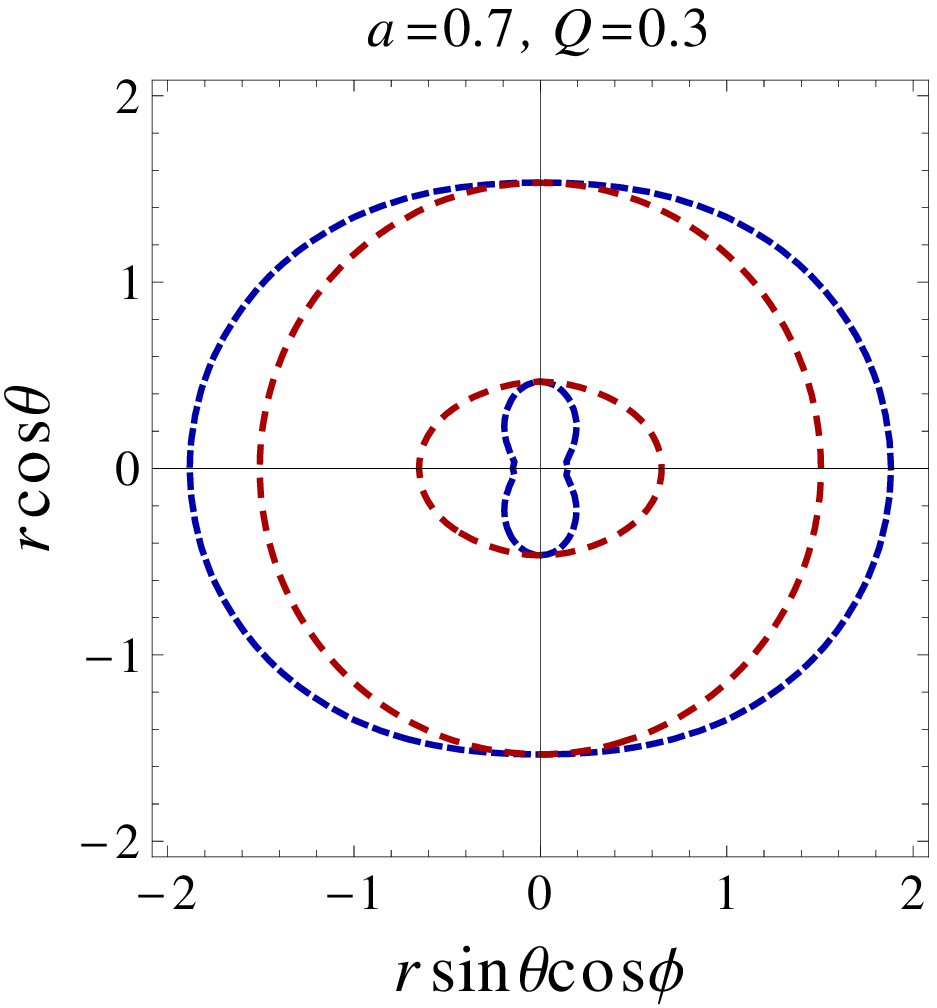}
\includegraphics[width=0.245\linewidth]{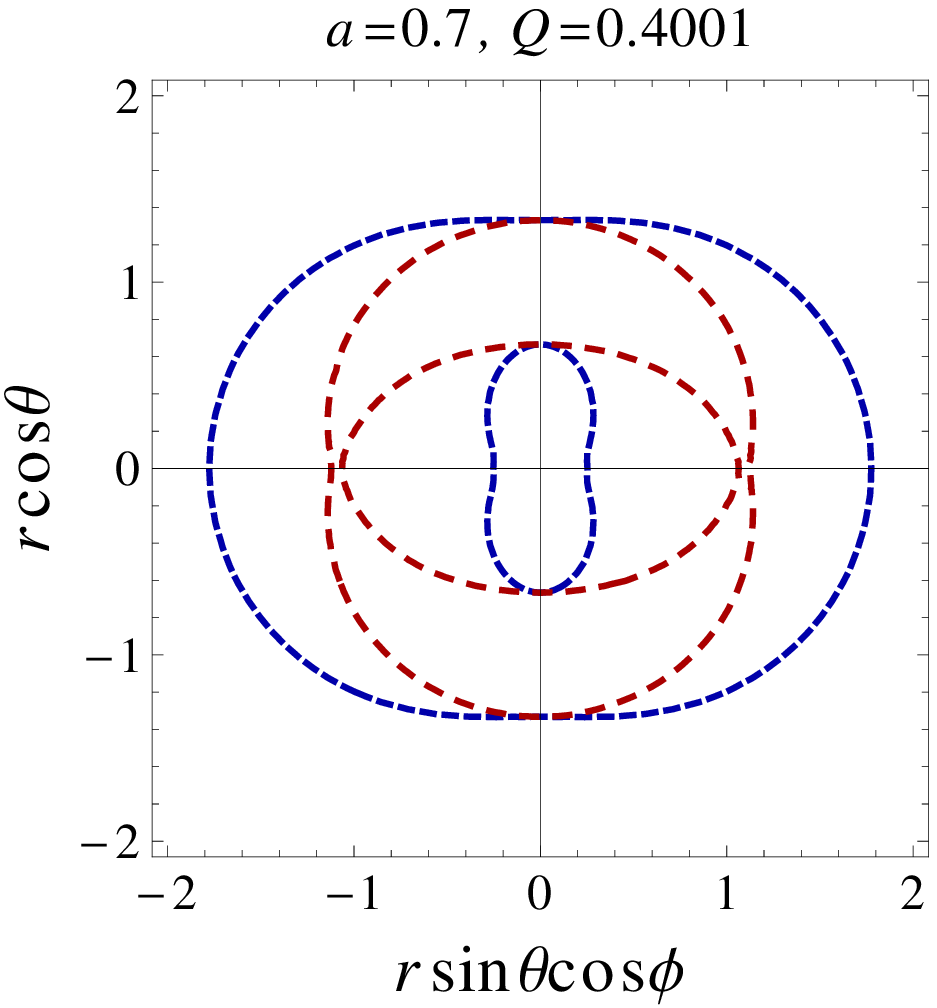}
\includegraphics[width=0.245\linewidth]{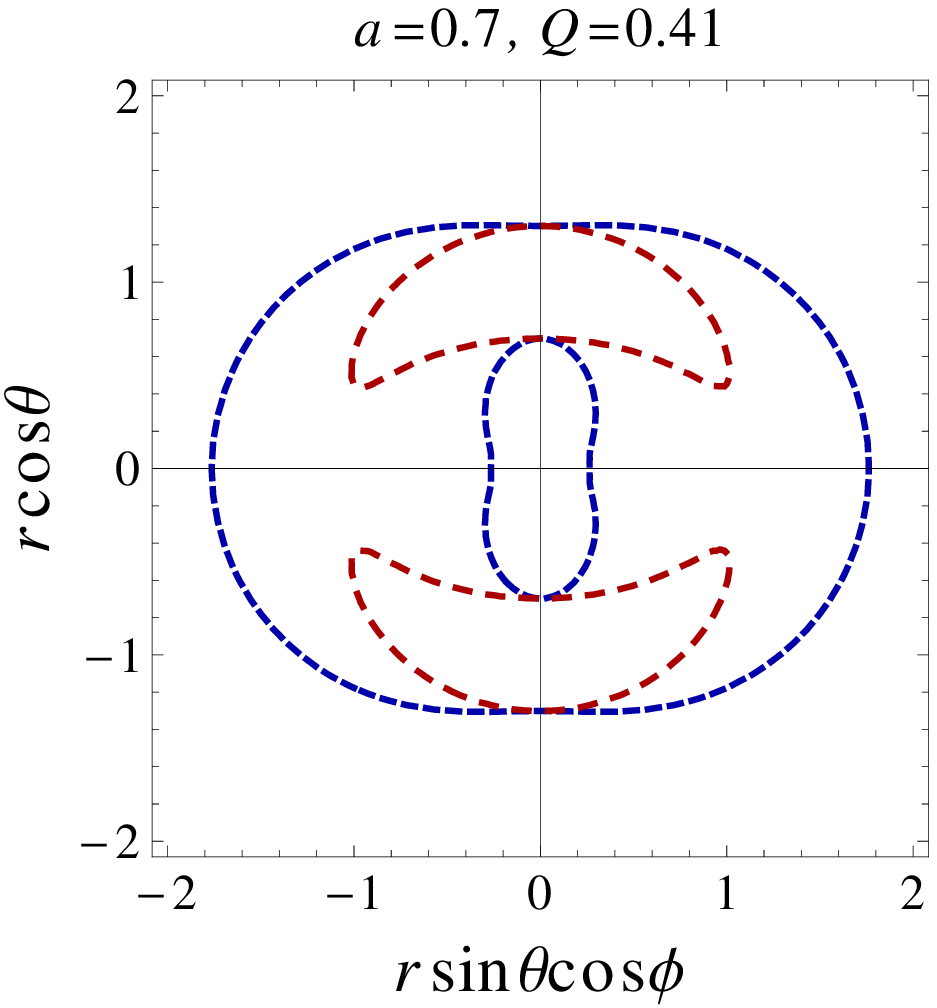}\\
\includegraphics[width=0.245\linewidth]{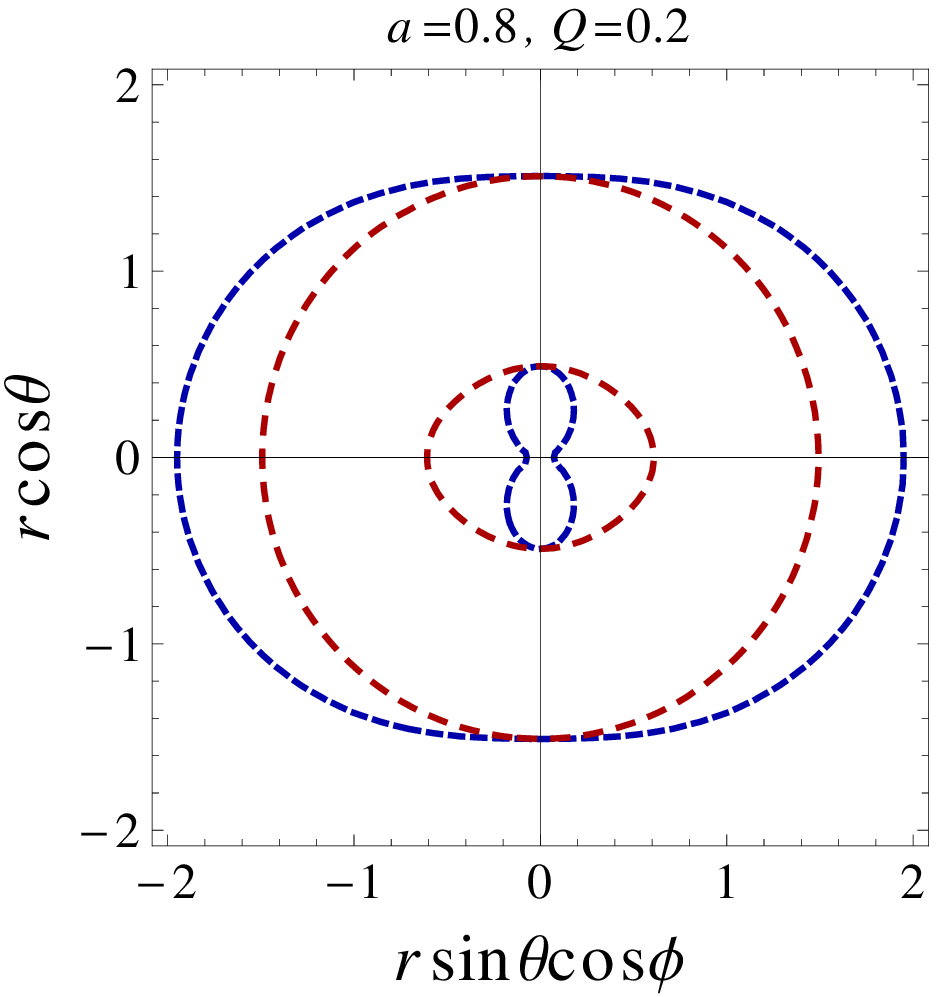}
\includegraphics[width=0.245\linewidth]{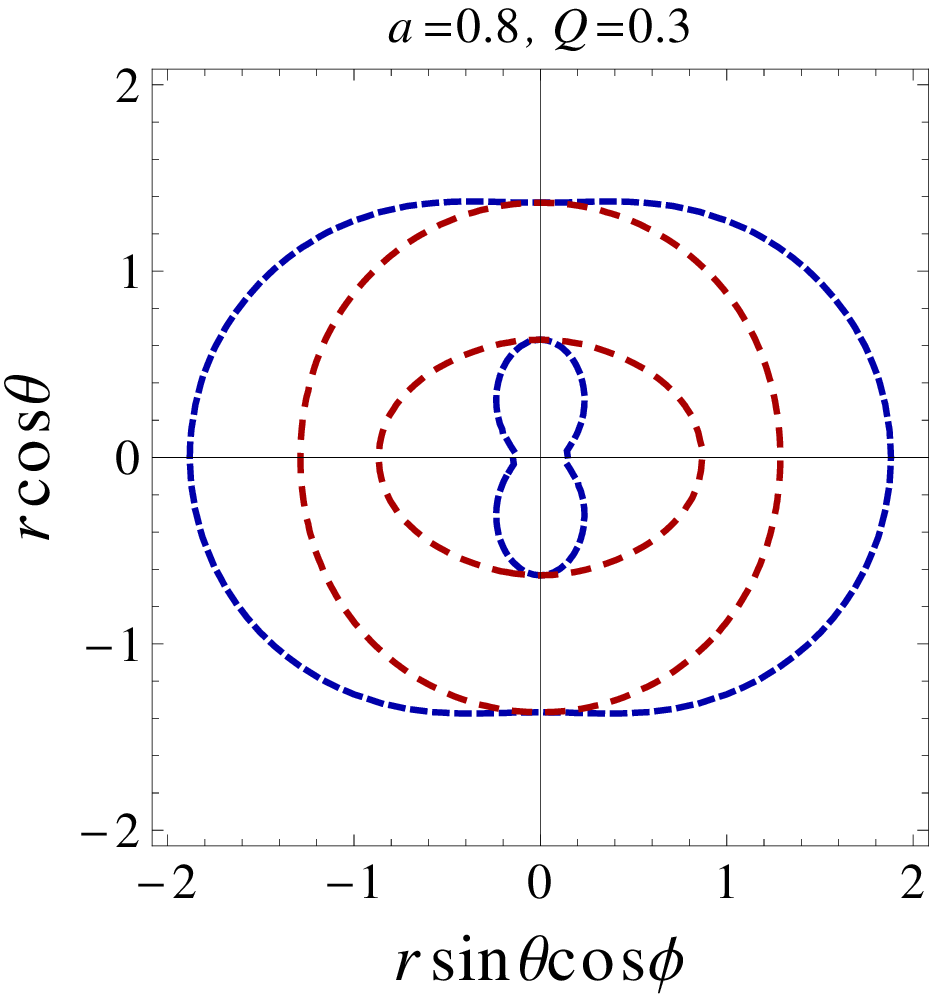}
\includegraphics[width=0.245\linewidth]{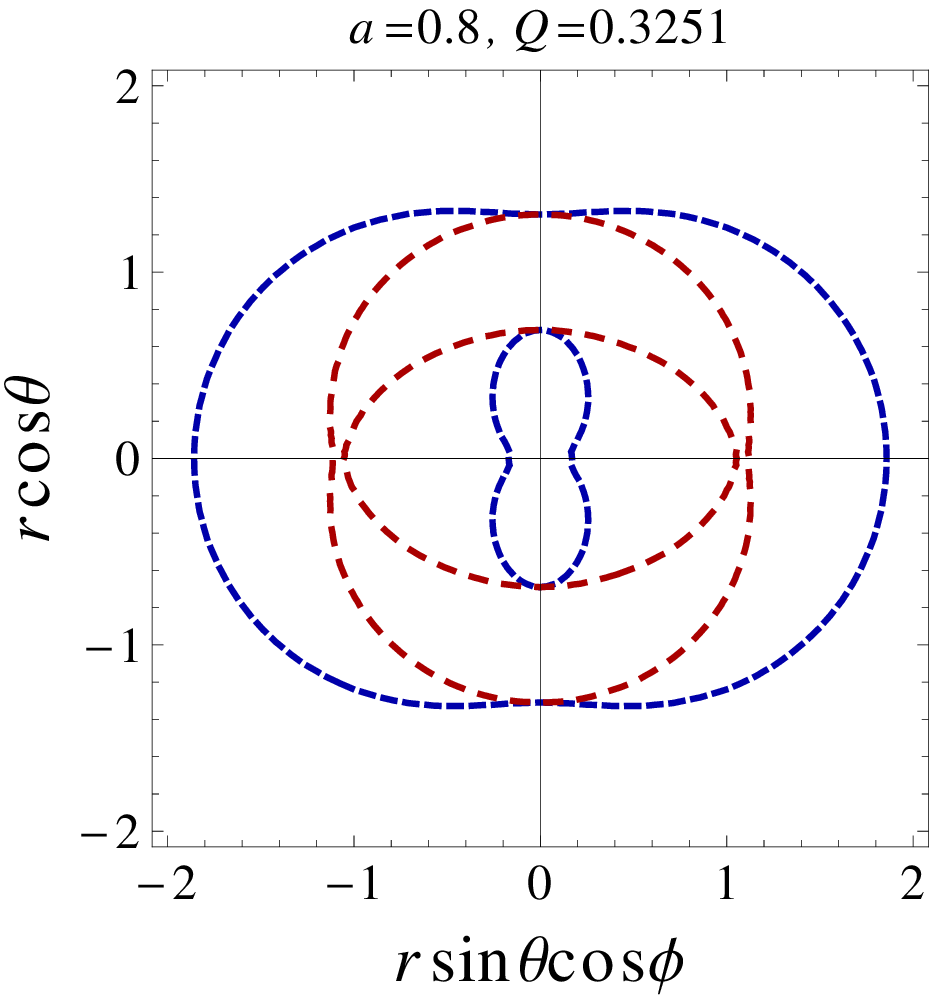}
\includegraphics[width=0.245\linewidth]{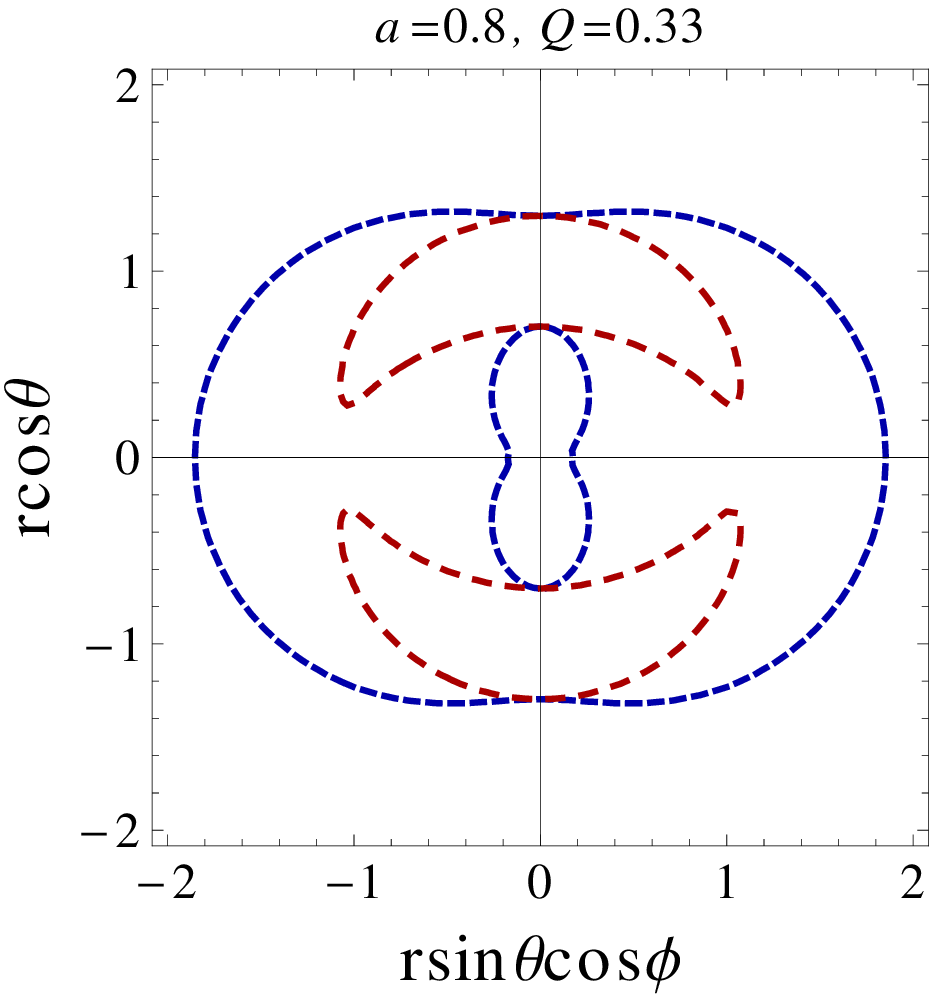}
   \end{tabular}
\caption{\label{figB} Plots showing the behavior of ergoregion in the xz-plane of rotating ABG black hole with charge $Q$ and for different values of rotation parameter $a$. The blue and red lines correspond to the static limit surface and horizons.}
\end{figure*}
How the charge parameter $Q$ affects the shape of the ergoregion is shown in Fig.~\ref{figB}, when compared with 
the Kerr black hole's ergoregion (cf. Fig.~\ref{figA}). The numerical solution of $g_{tt} =0$ and $g^{rr} =0$ are 
summarized in Table~\ref{Table A}. Fig.~\ref{figB} shows that the ergoregion area of the ABG black hole increases 
with increase in the value of $Q$ as well as $a$. The outer horizon and static limit surface coincides at the  
poles (Fig.~\ref{figB}) as in the case of Kerr black hole (Fig.~\ref{figA}). It should be pointed out that when 
$Q>Q_{C}$, for a given parameter $a$, the horizons get disconnected (cf. Fig.~\ref{figB}). The static limit 
surface becomes more oblate with increase in the value of $Q$.

\section{Equations of motion in the background of a rotating ABG black hole} \label{geqm}
\begin{figure*}
\begin{tabular}{c c c c}
\includegraphics[width=0.48\linewidth]{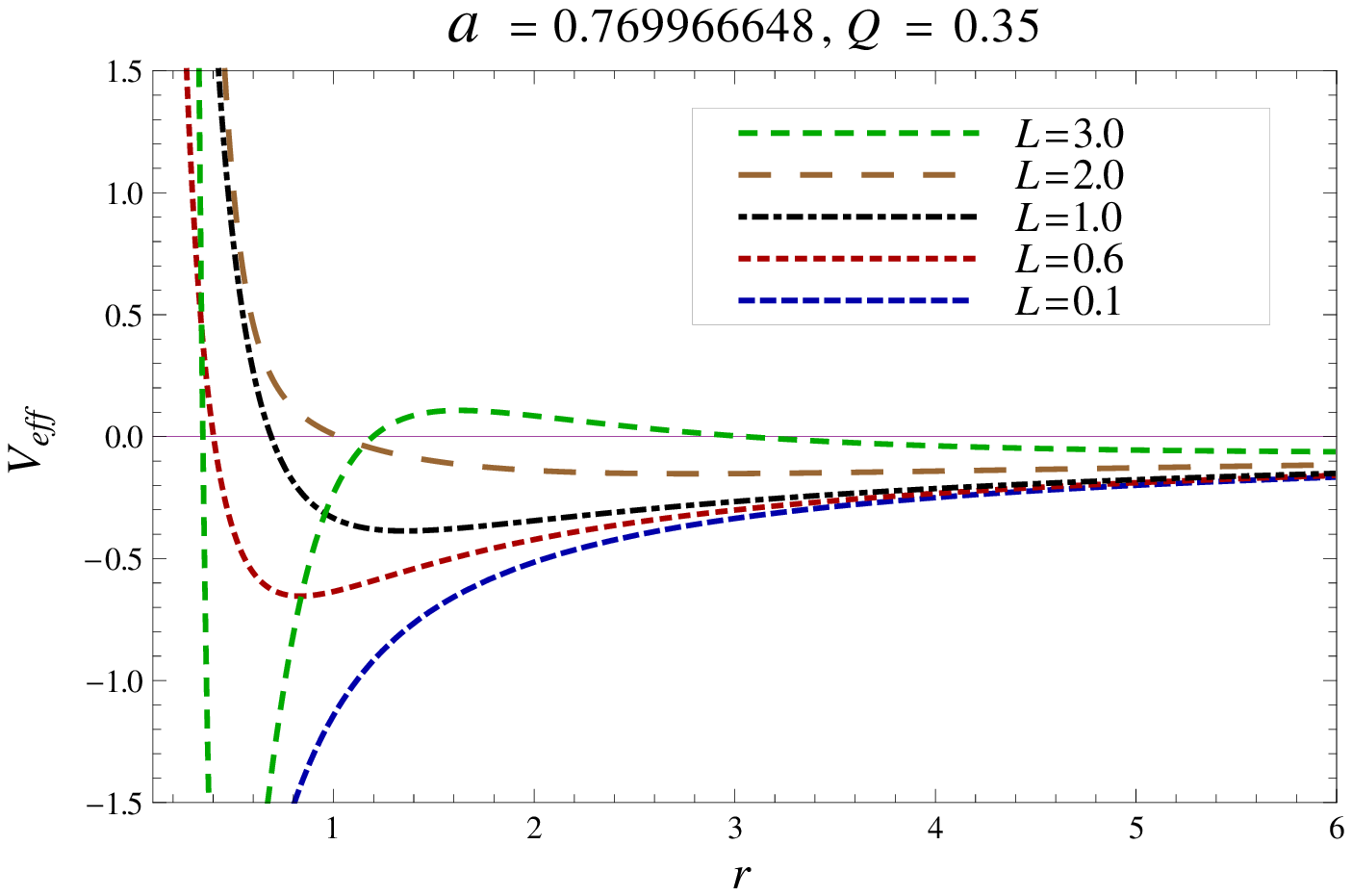}
\includegraphics[width=0.48\linewidth]{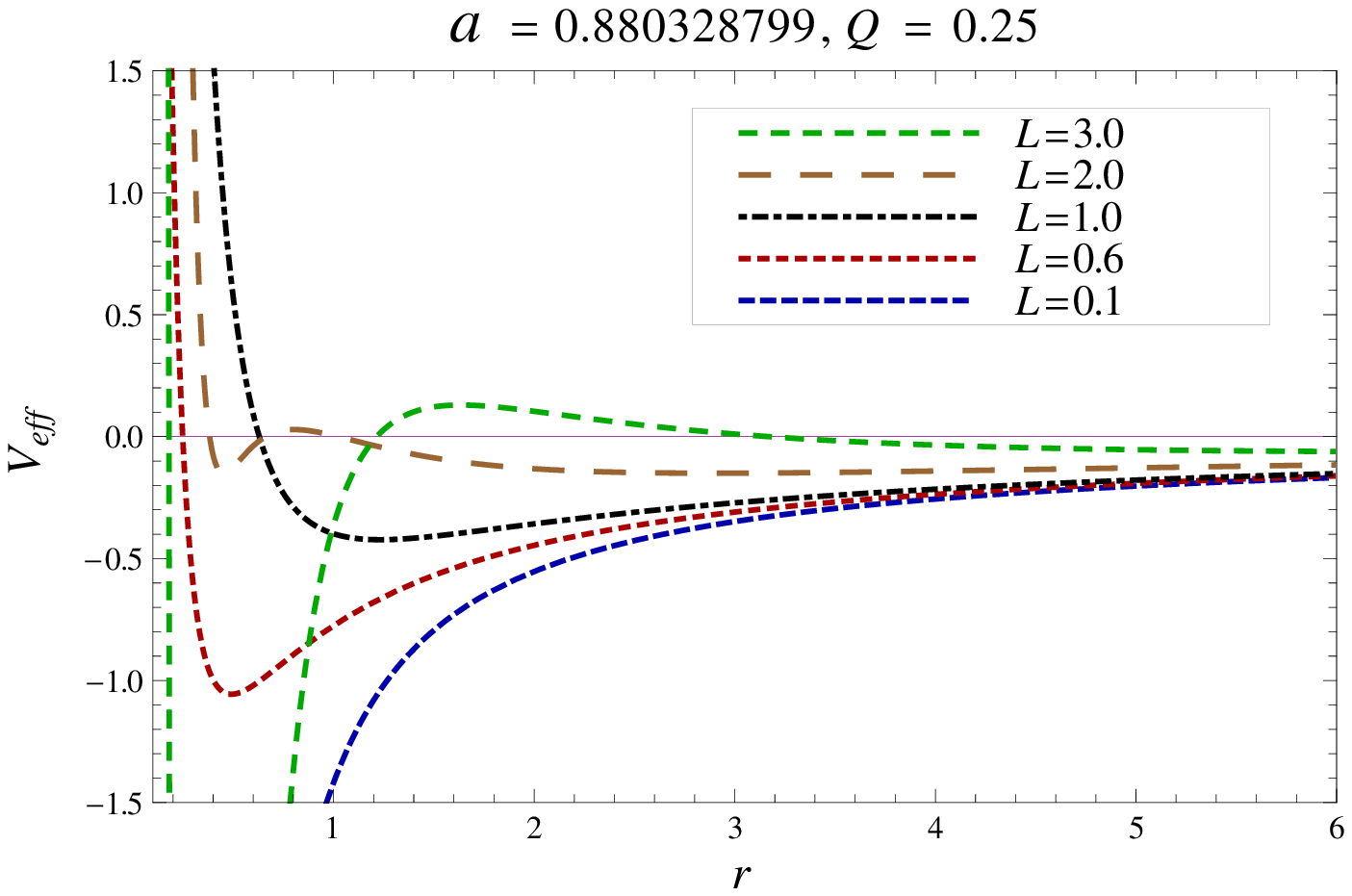}
\end{tabular}
\caption{\label{figPot} Plots showing the behavior of the effective potential $V_{eff}$ with radius $r$ for the different values of angular momentum.  }
\end{figure*}

Next, we calculate the equations of motion of the particle in the background of rotating ABG black hole, which are essential to study the collision of particles. Let us consider a particle of rest mass $m$ falling from rest at infinity on the equatorial plane ($\theta = \pi/2$) of a rotating ABG black hole. The motion of a particle is determined by the Lagrangian
\begin{eqnarray}
\mathcal{L} = \frac{1}{2} g_{\mu \nu} u^{\mu} u^{\nu},
\end{eqnarray}
where $u^{\mu}$ is the four-velocity. 
 The metric (\ref{metric}) has two conserved quantities, namely, energy $E$ and angular momentum $L$, respectively, correspond to the timelike Killing vector $\xi^a=(\partial/\partial t)^a$ and axial Killing vector $\chi^a=(\partial/\partial \phi)^a$, given \cite{6}
\begin{eqnarray}\label{el}
E = -g_{ab} \xi^a u^b,   \quad  L = g_{ab} \chi^a u^b,
\end{eqnarray}
which yields
\begin{eqnarray}
E = -g_{tt} u^t -g_{t \phi} u^{\phi},  \quad L = g_{t \phi} u^t +g_{\phi \phi} u^{\phi}. \nonumber
\end{eqnarray}
We obtain the four-velocities of the particle by solving the above equations,
\begin{eqnarray}\label{u^t}
 u^t &=& -\frac{1}{r^2} \Big[a (aE - L) - \frac{r^2 + a^2}{\Delta} \mathcal{P} \Big],
\end{eqnarray}
\begin{eqnarray}\label{u^Phi}
 u^{\phi} &=& -\frac{1}{r^2} \Big[(aE - L ) - \frac{a}{\Delta} \mathcal{P} \Big], 
\end{eqnarray}
and the normalization condition of the four-velocity $u_{\mu}u^{\mu}=-m^2$, yields
\begin{equation}\label{u^r}
 u^{r} = \pm \frac{1}{r^2} \sqrt{\mathcal{P}^2 -\Delta \left[m^2 r^2  + (L-a E)^2  \right]},
\end{equation}
where positive and negative sign correspond to the outgoing and incoming geodesics, respectively and 
$\mathcal{P} = (r^2 + a^2)E - aL $. The above equations have same expressions that of Kerr black hole, but now 
$\Delta$ is modified which includes charge $Q$. However, to obtain the trajectories of a particle, we need the 
angular momentum range of the particle and it can be calculated by using the circular orbit conditions,
\begin{eqnarray}\label{lim}
V_{eff}=0, \;\;\;\text{and}\;\;\; \frac{dV_{eff}}{dr}=0,
\end{eqnarray}
where $V_{eff}$ (effective potential) is
\begin{equation}
 V_{eff} =  -\frac{[(r^2 + a^2)E -La]^2 -\Delta [m^{2} r^2  + (L-a E)^2]}{2 r^4}.
\end{equation}
It determines the allowed and prohibited regions of the particle trajectory around the ABG black hole. 
  If $V_{eff} \leq 0$, then this is an allowed region and for $V_{eff} > 0$, the motion is prohibited. In Fig.~\ref{figPot}, we have shown the behaviour of $V_{eff}$ with radius $r$.
Since, the geodesics are timelike, i.e., $dt/d\tau \geq 0$, then using Eq.~(\ref{u^t}) we get
\begin{eqnarray}
\frac{1}{r^2}[-a(a E-L)+(r^2+a^2)\frac{\mathcal{P}}{\Delta}]\geq 0,
\end{eqnarray}
at horizon, the above condition on simplification leads to
\begin{eqnarray}
E-\Omega_H L \geq 0, \quad \Omega_H = \frac{a}{(r^{EH}_+)^2+a^2}.
\end{eqnarray}

\begin{widetext}
\begin{table}
\caption{Table for the range of the angular momentum in extremal cases of a rotating ABG black hole ($M=1$). }
\begin{center}\label{Table B}
\begin{tabular}{| c | c | c | c |  c|}
\hline
 $ Q $ & $ a = a_{ex} $ & $ r=r_{ex} $  & $ L_{1} $ & $ L_{2} $  \\
\hline  
0     &  1           &  1      &  2.0     &  -4.82842  \\
0.15  &  0.955870457  &  1.02859  &  2.06269 &  -4.78476  \\ 
0.25  &  0.880328799  &  1.06201  &  2.16152  &  -4.70709   \\
0.35  &  0.769966648  &  1.08805  &  2.30749  &  -4.58729  \\
0.45  &  0.621056295  &  1.09518  &  2.55058  &  -4.41379  \\
\hline   
\end{tabular}
\end{center}
\end{table}
\end{widetext}

The critical angular momentum of the particle is defined by $L_C = E / \Omega_H$, where $\Omega_H$ is the angular 
velocity  at the horizon of the ABG black hole. We have shown the values of critical angular momentum for the 
extremal cases of rotating ABG black hole in Table~\ref{Table B}. It is known that the particle trajectory 
depends on the values of angular momentum which is different for different values of angular momentum. If angular 
momentum of the particle is larger than the critical angular momentum, i.e., $L>L_C$, then the particle will never 
fall into the black hole. When the angular momentum of particle is smaller then the critical angular momentum 
$L<L_C$, then the particle will always fall into the black hole. Moreover, when the particle's angular momentum 
is equal to the critical angular momentum $L=L_C$, then the particle reaches up to the horizon and collision of 
two particle takes place.

\section{Center-of-mass energy of two particles in the rotating ABG spacetime} \label{cme}
\begin{figure*}
\includegraphics[width=0.48\linewidth]{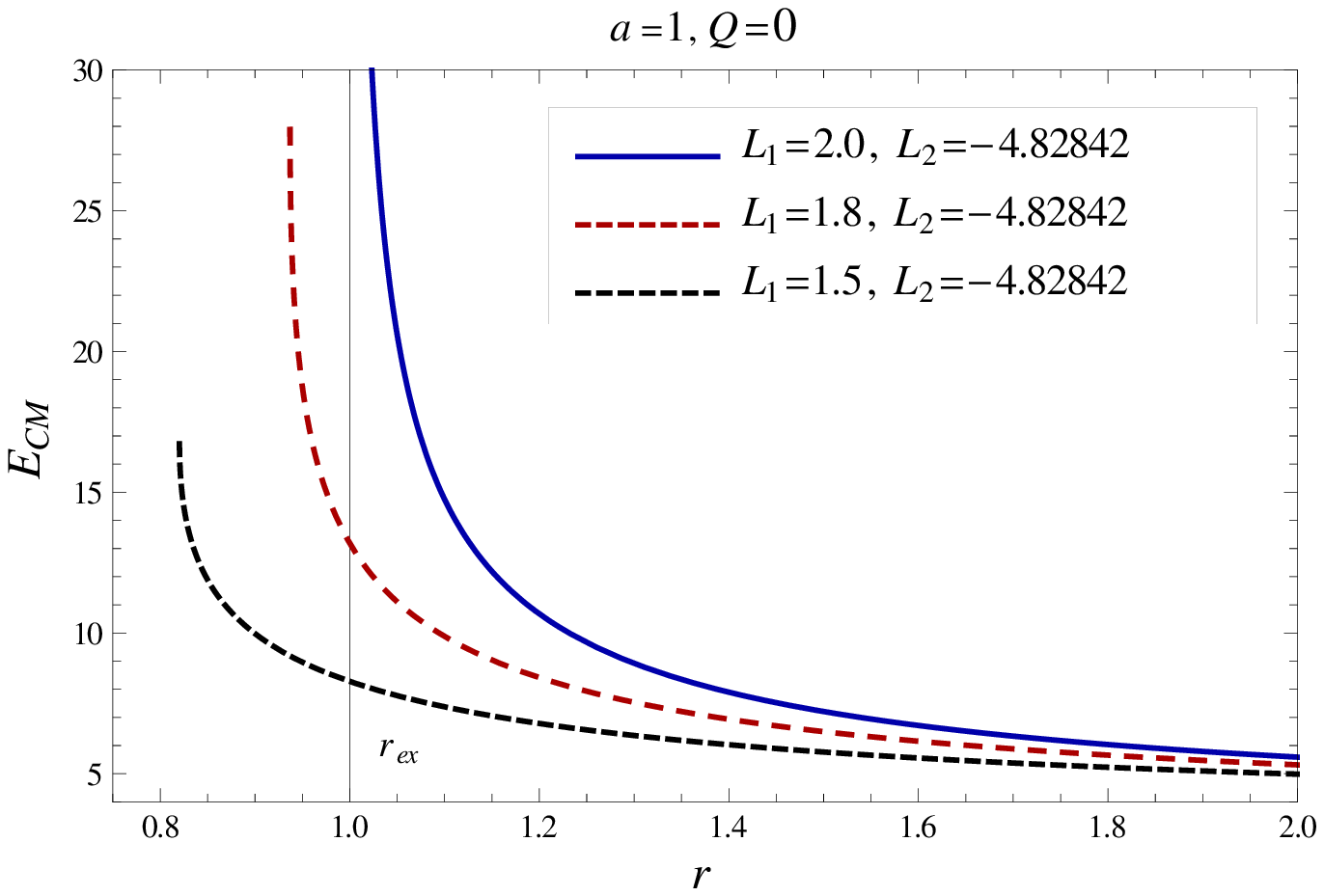}
\includegraphics[width=0.48\linewidth]{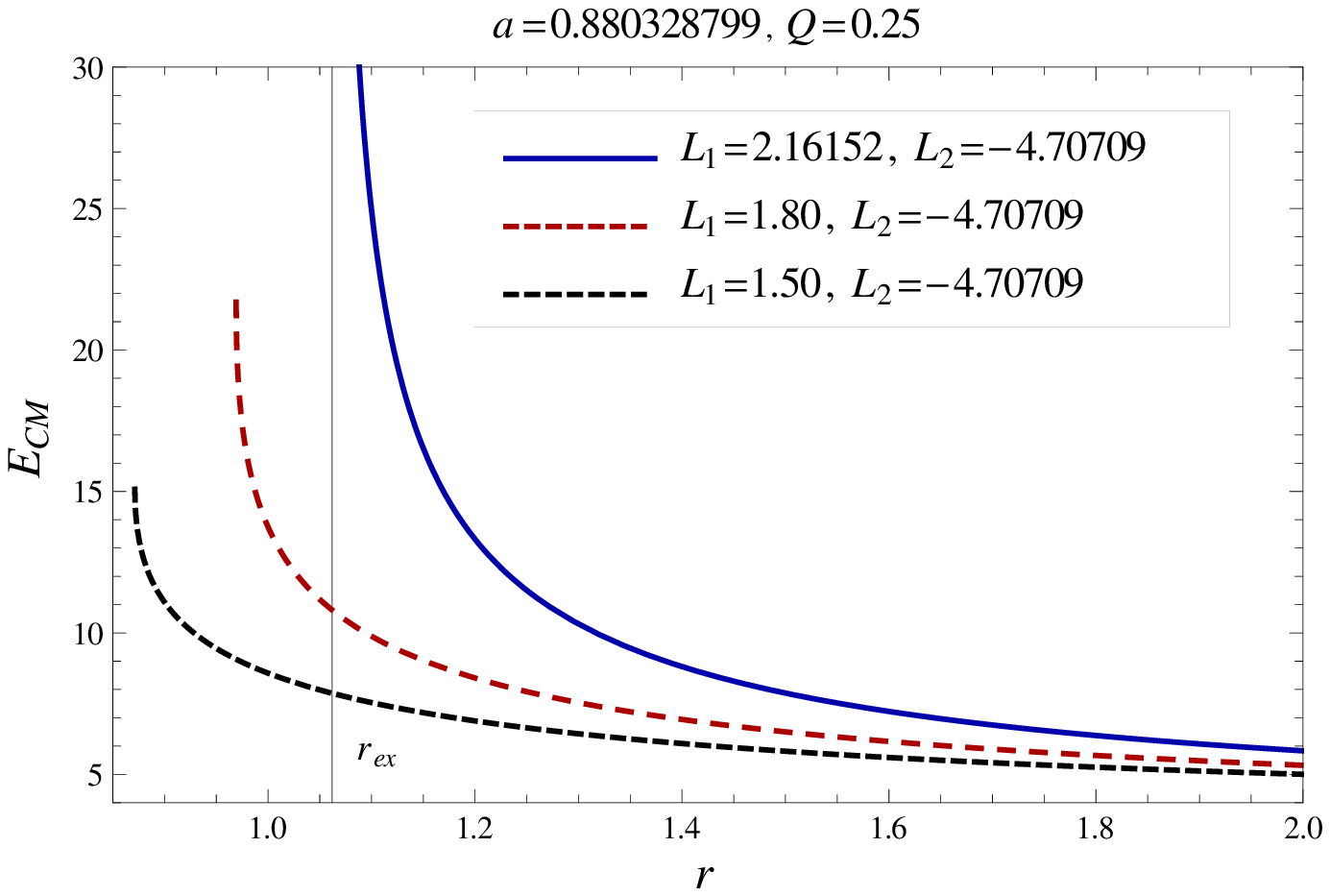}
\includegraphics[width=0.48\linewidth]{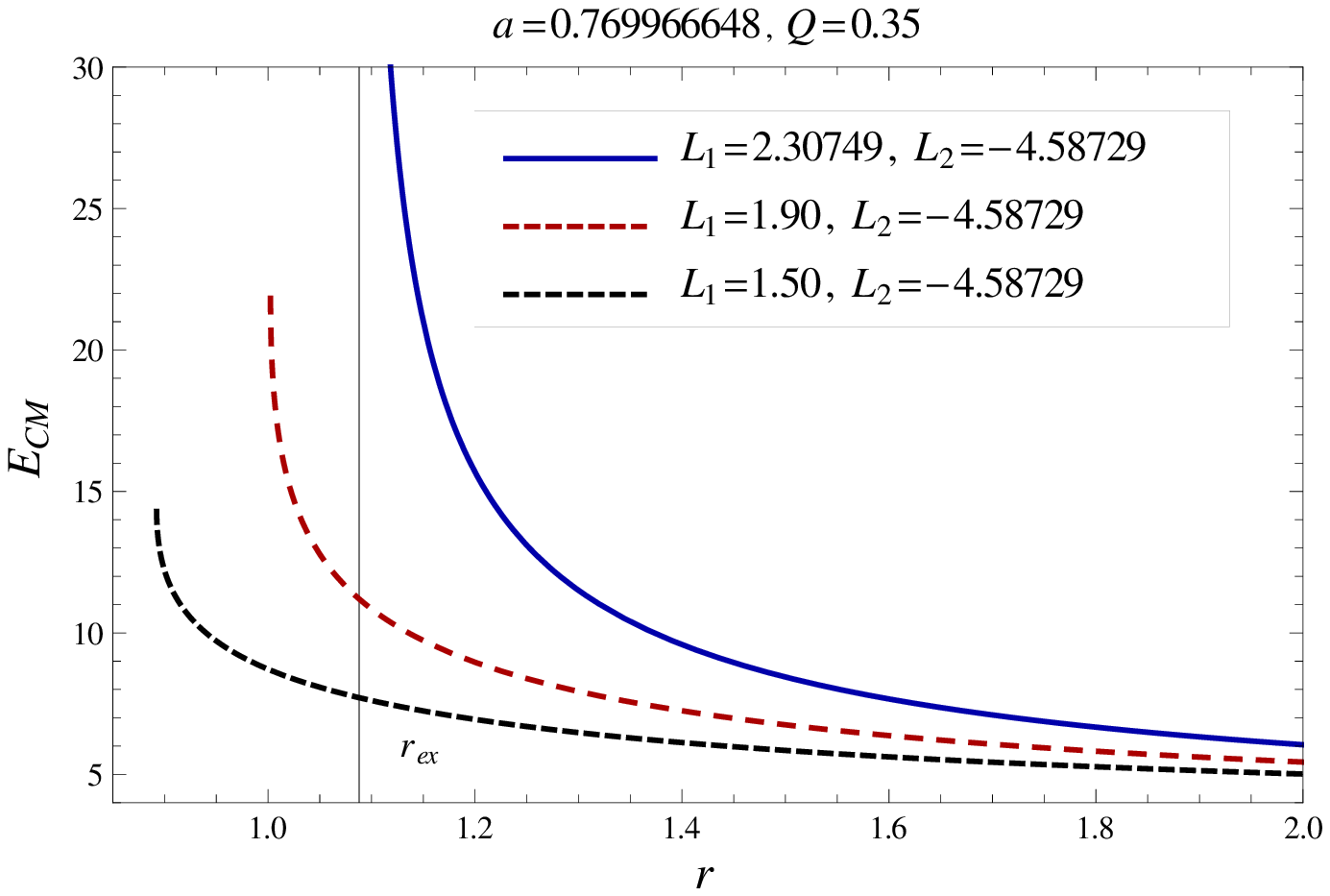}
\includegraphics[width=0.48\linewidth]{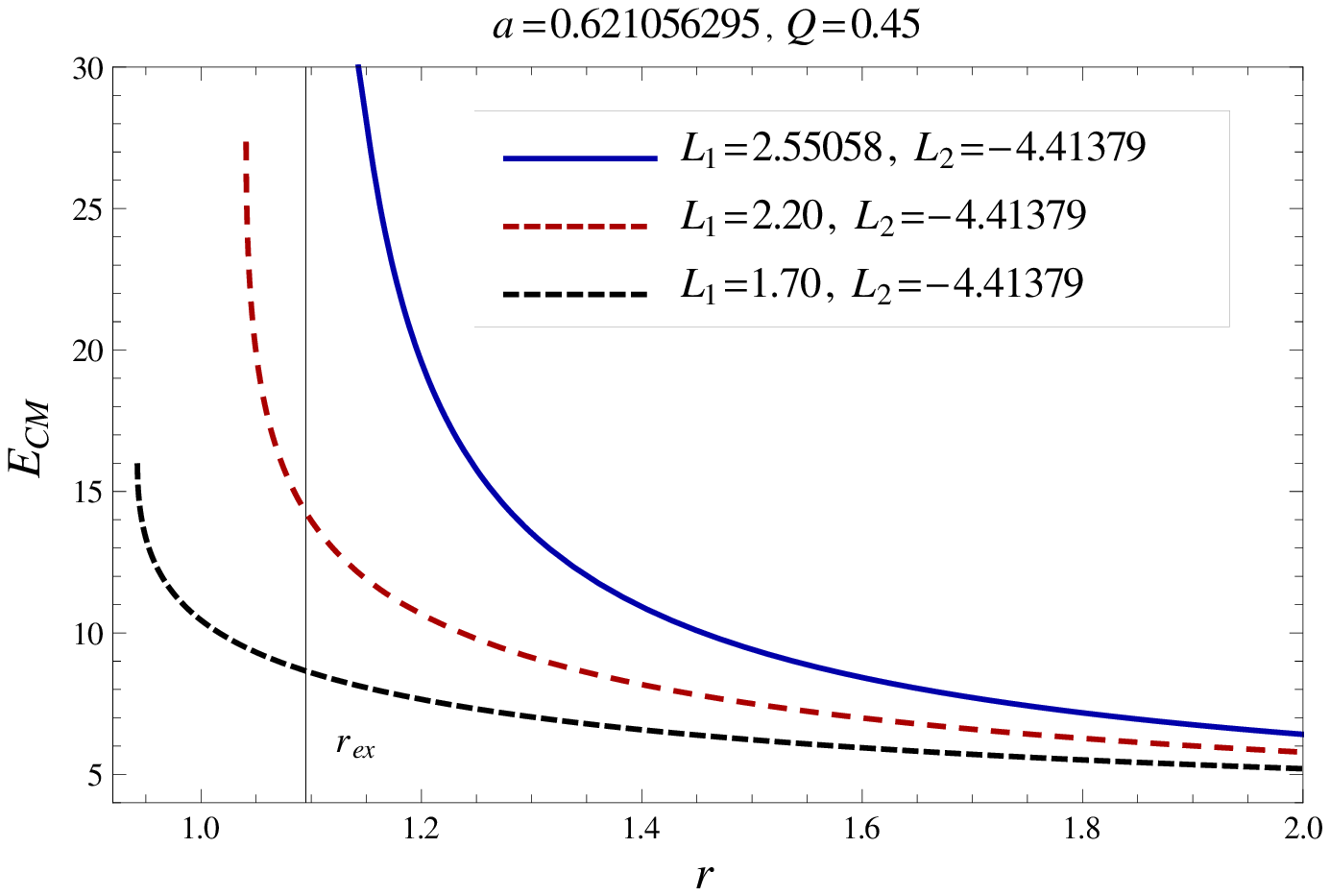}
\caption{\label{fig3} Plots showing the behavior of $E_{CM}$ vs $r$ for an extremal rotating ABG black hole 
with colliding particles masses $m_{1}=1, m_{2}=2$. $r_{ex}$ corresponds to the extreme value of the horizon.}
\end{figure*}

After calculating the equations of motion of a particle, and checked the conditions for the particle to reach up to the horizon, now we are in a position to calculate the center-of-mass energy ($E_{CM}$) of two different mass particles. Let us consider the two particles with rest masses $m_{1}$ and $ m_{2}$ ($m_{1}\neq m_{2}$)  moving in the equatorial plane ($\theta = \pi/2$) of rotating ABG black hole. These particles are coming from rest at infinity towards the black hole and collide in the vicinity of an event horizon. The collision takes place in the center-of-mass frame. The two particles which are involved in the collision have different angular momentum, i.e., $L_{1}$, $L_{2}$ and energy $E_1$, $E_2$. The four-momentum of the $i^{th}$ particle is given by \cite{Banados:2009pr}
\begin{equation}
p^{\mu}_{i}=m_{i} u^{\mu}_{i},
\end{equation}
where  $i=1,2$ and $m_{i}$ and $u^{\mu}_{i}$ are corresponding to the rest mass and four-velocity of the $i^{th}$ particle. The total four-momentum of two colliding particles is
\begin{equation}
P^{\mu}_{t} = P^{\mu}_{(1)} + P^{\mu}_{(2)}.
\end{equation}
Hence, the $E_{CM}$ for the collision of two different mass particles is given by \cite{6}
\begin{eqnarray}\label{pmu}
E^{2}_{CM} = -P^{\mu}_{t} P_{t \mu} = -(P^{\mu}_{(1)} + P^{\mu}_{(2)})(P_{(1)\mu} + P_{(2)\mu}) \nonumber\\ 
= -\left(m_1 u^{\mu}_{(1)}+ m_2 u^{\mu}_{(2)}\right)\left(m_1 u_{(1)\mu}+m_2 u_{(2)\mu} \right). 
\end{eqnarray}
After simplifying and using the condition $u^{\mu}_{(i)} u_{(i) \mu} =-1$, in Eq.~(\ref{pmu}), we obtain the 
formula for $E_{CM}$ is 
\begin{equation}\label{formula}
\frac{E_{CM}^2}{2 m_{1} m_{2}} = 1+\frac{(m_{1} - m_{2})^2}{2 m_{1} m_{2}} 
- g_{\mu \nu} u^{\mu}_{(1)} u^{\nu}_{(2)}.
\end{equation}
By substituting the values of $g_{\mu \nu}$,  $u^{\mu}_{(1)}$ and $u^{\nu}_{(2)}$ from Eqs.~(\ref{metric}), 
(\ref{u^t}), (\ref{u^Phi}), and (\ref{u^r}) into Eq.~(\ref{formula}), therefore the $E_{CM}$ of rotating ABG 
black hole takes the following form: 
\begin{eqnarray}\label{ecm}
\frac{E_{CM}^2}{2 m_{1} m_{2}} &=& \frac{(m_{1}-m_{2})^{2}}{2 m_{1} m_{2}}+ \frac{1}{r^2(r^2 f(r) + a^2)} 
\Big[ a (f(r)-1)(L_{1} E_{2} + L_{2} E_{1})r^2  
\nonumber\\ && 
-a^2((f(r)-2)E_1E_2-1)r^2 - L_{1} L_{2} f(r)r^2+ (E_1E_2+ f(r))r^4
\nonumber \\ && 
-\sqrt{-r^2 [\left((f(r)-2)E^2_{1}+m_1^2\right)a^2-2a(f(r)-1)E_{1}L_{1}-r^2E_1^{2}+f(r)(L^2_{1}+r^2 m_1^2)]}
\nonumber \\ &&
\times \sqrt{-r^2 [\left((f(r)-2)E^2_{2}+m_2^2\right)a^2-2a(f(r)-1)E_{2}L_{2}-r^2E_2^{2}+f(r)(L^2_{2}+r^2 m_2^2)]}
\Big].\nonumber \\
\end{eqnarray}
We have considered the motion of the particles in equatorial plane ($\theta = \pi/2$), where $f(r, \theta)= f(r)$.
It is clear from Eq.~(\ref{ecm}), that $E_{CM}$ depends on charge $Q$ as well as rotation parameter $a$. If we 
change the value of charge $Q$ or rotation parameter $a$, then the $E_{CM}$ must be change. We analyze  
numerically the behavior of $E_{CM}$ with radius $r$. We plot Eq.~(\ref{ecm}) for different combinations of 
rotation parameter $a$, electric charge $Q$, angular momentum $L_{1}$ and $L_{2}$, and colliding particle masses 
$m_1$ and $m_2$. We observe from the Fig.~\ref{fig3} that the $E_{CM}$ would be very large if one of the colliding 
particle has critical value of angular momentum and collision takes place in the vicinity of the event horizon. 
Keep in mind that Eq.~(\ref{ecm}) is valid for two different massive particles. 

If we consider the limit $Q \rightarrow 0$ , then it follows from Eq.~(\ref{ecm}) that
%\begin{widetext}
\begin{eqnarray}\label{ecm2}
\frac{E_{CM}^2}{2 m_{1} m_{2}}(Q \rightarrow 0) &=& \frac{(m_{1}- m_{2})^{2}}{2 m_{1} m_{2}}+\frac{1}{r(r^2-2 M r+a^2)} \Big[a^2((2M+r)E_1 E_2 +r) \nonumber\\ &&-2 a M(L_1 E_2+L_2 E_1)-L_1 L_2 (-2M+r)+(-2M+r(1+E_1 E_2))r^2
\nonumber\\ 
&& -\sqrt{r(r^2+a^2)(E^2_{1}-m_1^2)+2M(a E_{1}-L_{1})^2 - L_{1}^2 r + 2M r^2 m_1^2} \nonumber \\ 
&& \times \sqrt{r(r^2+a^2)(E^2_{2}-m_2^2)+2M(a E_{2}-L_{2})^2 - L_{2}^2r + 2 M r^2 m_2^2}\Big],
\end{eqnarray}
%\end{widetext}
which is the center-of-mass energy when two particles of different masses collide near the Kerr black hole 
\cite{Harada:2011xz}. It can be analyze that when $m_{1}=m_{2}=m_0$ and $E_1=E_2=E=1$, then Eq.~(\ref{ecm}) 
reduces to
%\begin{widetext}
\begin{eqnarray}\label{ecm1}
\frac{E_{CM}^2}{2 m_{0}^2} &=&\frac{1}{r^2(r^2 f(r)+a^2)}\Big[a(f(r)-1) (L_{1}+L_{2})r^2 -a^2 (f(r)-3)r^2 - L_{1}L_{2}f(r)r^2  \nonumber\\ && 
+ (1+f(r))r^4 - \sqrt{-r^2[(f(r)-1)a^2-2a(f(r)-1)L_{1}-r^2+f(r)(L_{1}^{2}+r^2)]} \nonumber\\ && 
\times \sqrt{-r^2[(f(r)-1)a^2-2a(f(r)-1)L_{1}-r^2+f(r)(L_{1}^{2}+r^2)]}\Big],
\end{eqnarray}
%\end{widetext}
which is similar to the $E_{CM}$ of rotating ABG black hole for two equal mass particles \cite{Ghosh:2014mea}. 
Again, if $Q \rightarrow 0$ and $m_{1}=m_{2}=m_0$, and $E_1=E_2=E=1$, then Eq.~(\ref{ecm}) transform into
the $E_{CM}$ of Kerr black hole \cite{Banados:2009pr}:
\begin{widetext}
\begin{eqnarray}\label{ecm3}
{E_{CM}^2}(Q \rightarrow 0) &=&\frac{2 m_0^{2}}{r(r^2 -2r +a^2)}\Big[2 a^2 (1+r)-2a (L_{1}+L_{2})-L_{1} L_{2}(-2+r) +2(-1+r)r^2 \nonumber\\ 
&& -\sqrt{2(a-L_{1})^2- L_{1}^2 r+2 r^2} \sqrt{2(a-L_{2})^2- L_{2}^2 r+2 r^2} \Big].
\end{eqnarray}
\end{widetext}
Hence, we can say that the $E_{CM}$ of two different mass particles in the background of rotating ABG black hole 
spacetime is the generalization of $E_{CM}$ of Kerr black hole. Now we discuss the $E_{CM}$ of two colliding 
particles for extremal and nonextremal black hole.
\begin{figure*}
\includegraphics[width=0.48\linewidth]{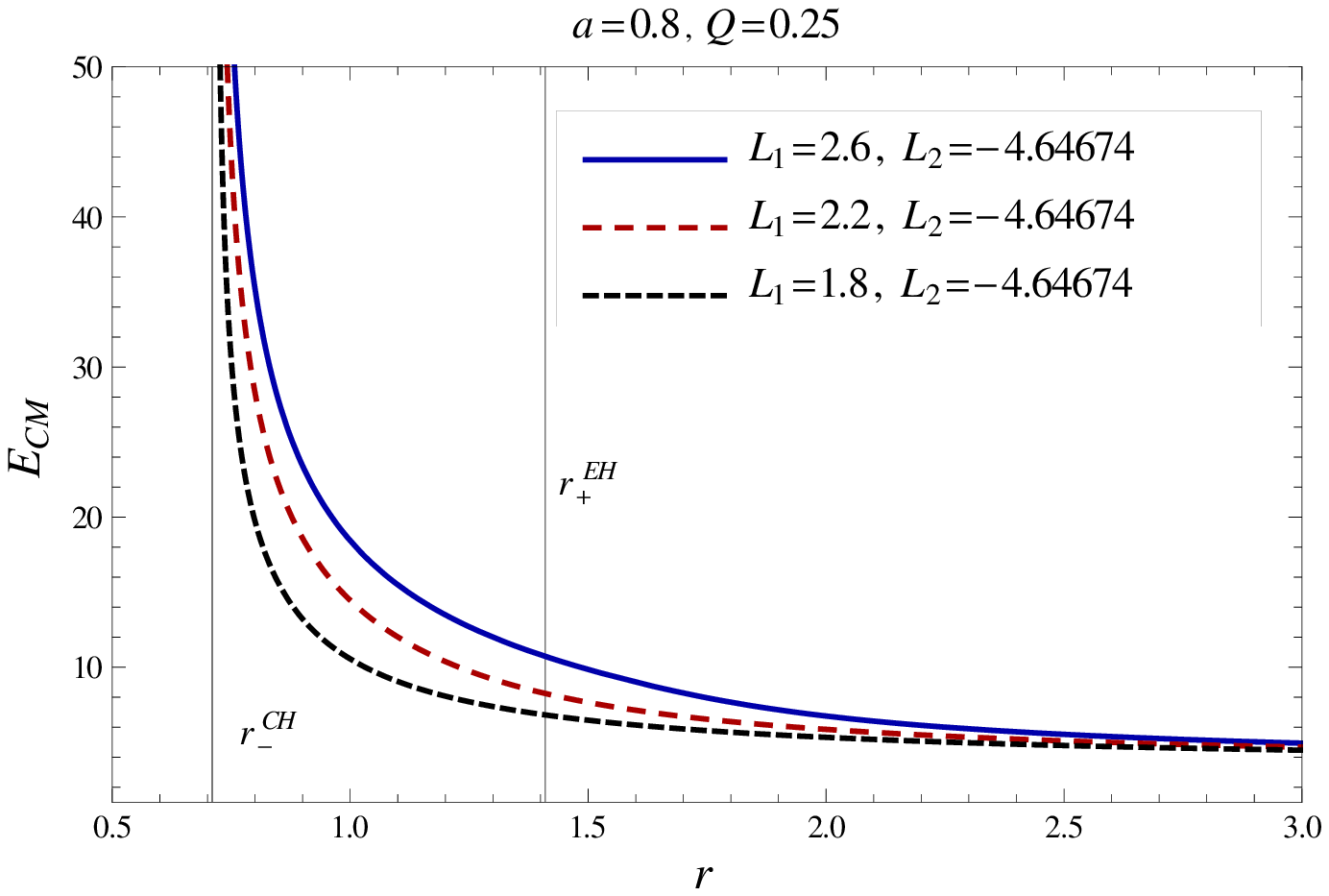}
\includegraphics[width=0.48\linewidth]{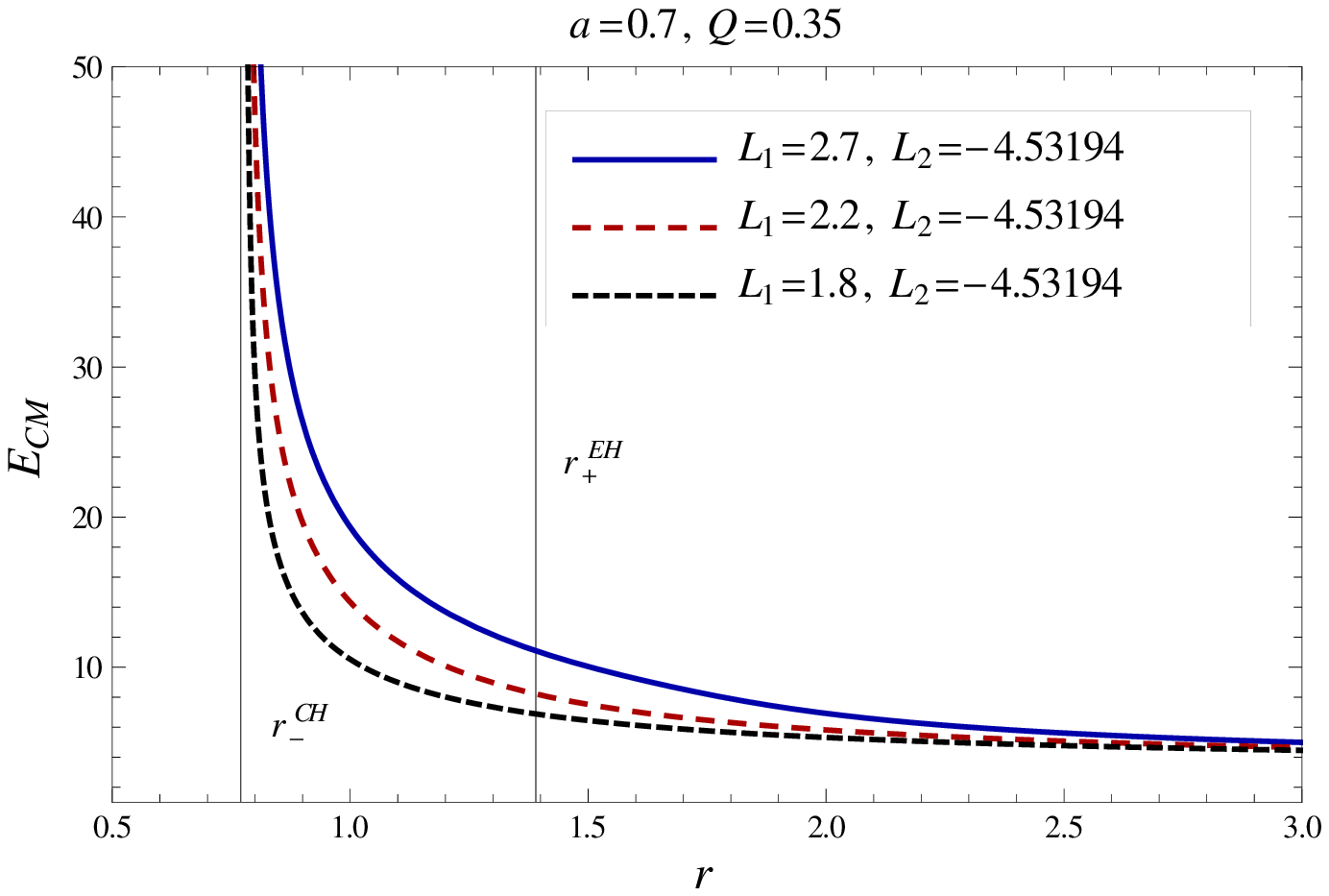}
\caption{\label{fig4} Plots showing the behavior of $E_{CM}$ vs $r$ for nonextremal rotating ABG black hole with colliding particles masses $m_{1}=1, m_{2}=2$. Vertical lines correspond to the outer horizon ($r^{EH}_{+}$) and inner horizon ($r^{EH}_{-}$).}
\end{figure*}
\begin{figure*}
\includegraphics[width=0.48\linewidth]{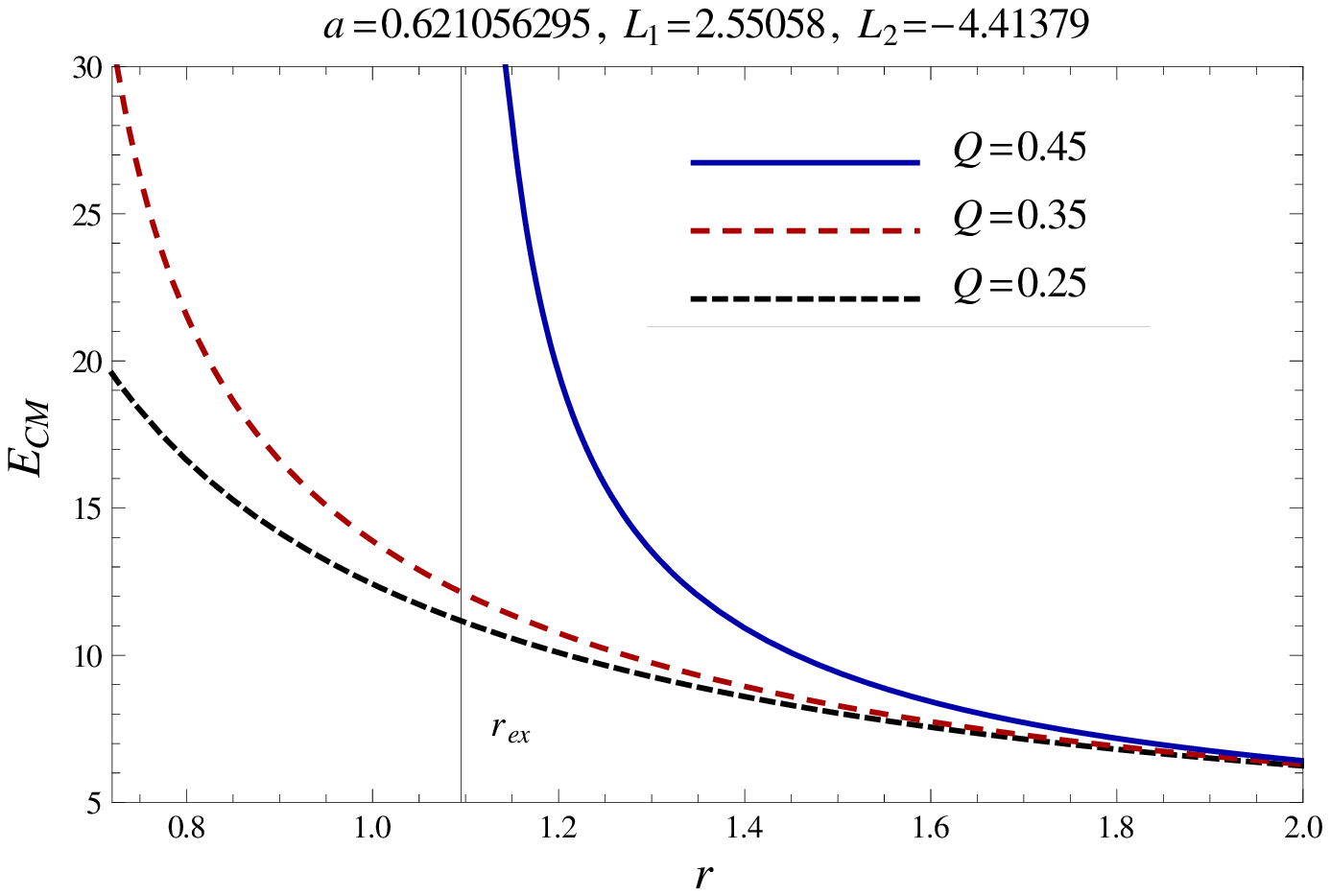}
\includegraphics[width=0.48\linewidth]{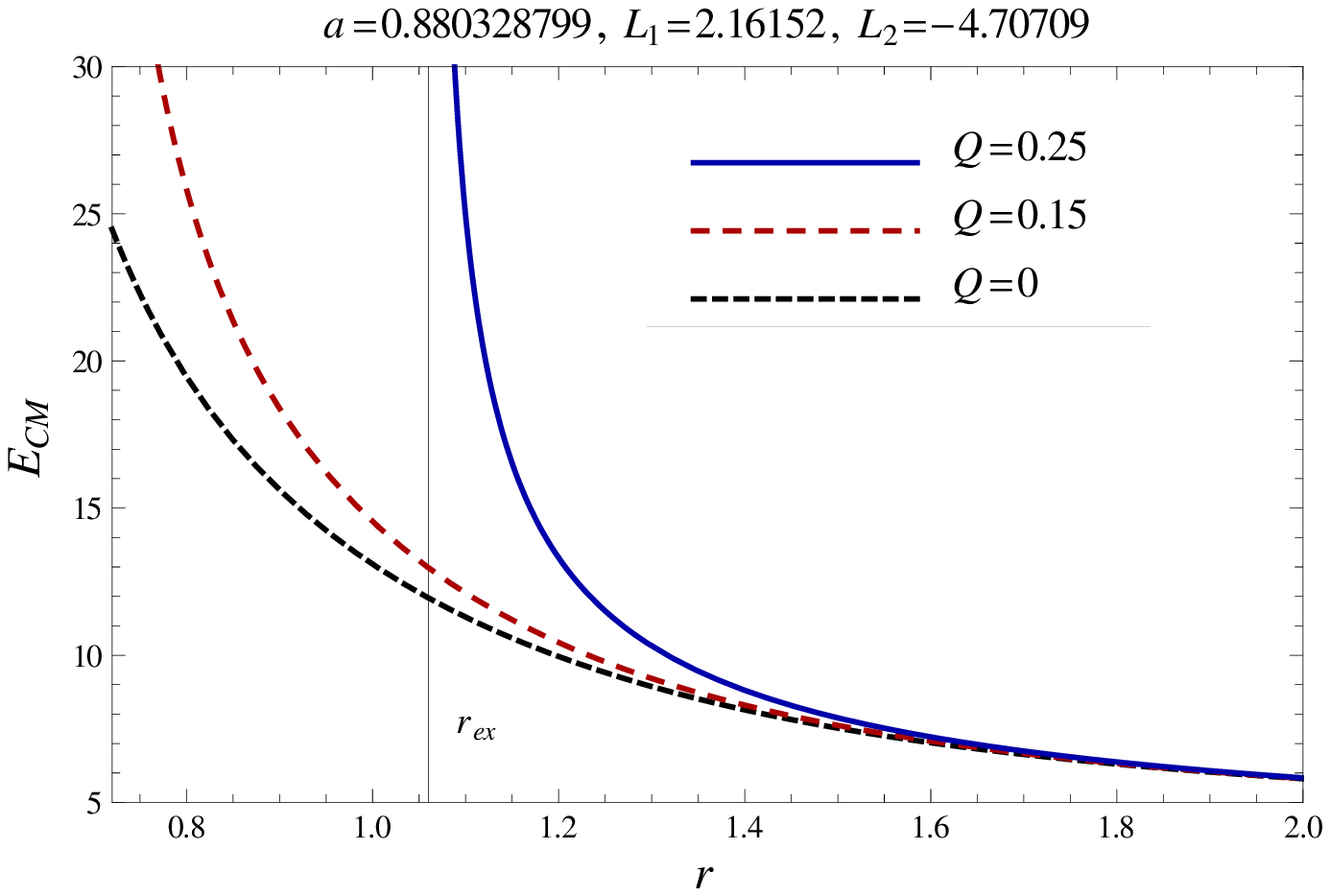}
\caption{\label{fig5} Plots showing the behavior of $E_{CM}$ vs $r$ for different values of $Q$ for a  rotating ABG black hole with colliding particles masses $m_{1}=1, m_{2}=2$. $r=r_{ex}$ correspond to the horizon for the extremal cases.}
\end{figure*}

We are interested in the near horizon collision of two different mass particles in case of extremal black hole, 
i.e., $r \rightarrow r_{ex}$. We analyze the behavior of $E_{CM}$ with radius $r$, and plot it in Fig.~\ref{fig3}, 
which show that the $E_{CM}$ due to the collision of two different mass particles is infinite when one of the 
particles have a critical angular momentum (for a suitable choice of rotation parameter $a$ and charge $Q$). From 
the graphical representation of $E_{CM}$, we have seen that $E_{CM}$ is infinite for 
$L_1=2.0,2.16152,2.30749,2.55058$ corresponding to $Q=0.0,0.25,0.35,0.45$ and for all other values of angular 
momentum, it remains finite. 

After studied the near horizon collision in extremal case, now we extend it for nonextremal black holes. 
We can see the behavior of $E_{CM}$ vs $r$ for nonextremal black hole from Fig.~\ref{fig4} for different 
charge $Q$ and different mass. From Fig.~\ref{fig4}, we can conclude that $E_{CM}$ would be infinite, if 
collision takes place at the inner horizon  of the nonextremal black hole. If the particles collide at the 
outer horizon  of the black hole, then the $E_{CM}$ will remain finite. So, for getting a very high energy, 
particle collision should happen at the inner horizon. Also, in  Fig.~\ref{fig5}, we have shown the effect 
of $Q$ on the $E_{CM}$.
 
\section{Conclusion} 
\label{con}
The gravitational collapse of sufficiently massive star leads to the formation of spacetime singularities is a 
quite common phenomenon in general relativity as predicated by famous singularity theorems \cite{he}. However, 
there is a belief that these spacetime singularities do not exist in Nature, but that they are artefact of the 
classical general relativity. On the other hand, the cosmic censorship conjecture asserts that these singularities 
can not be seen by an external observer \cite{rp}. However, the conjecture does not ruled out the possibility of 
regular black holes. In this paper, we consider the rotating ABG black holes, which can be written in Kerr-like 
form in Boyer-Lindquist coordinates with mass $M$ and has an additional parameter charge ($Q$) that measures 
potential deviations from the Kerr metric and includes the Kerr metric as the special case in the absence of the 
charge ($Q=0$). This special solution is a solution of general relativity coupled to nonlinear electrodynamics 
that is of Petrov type D and it is singularity free. We have studied in detailed the horizon properties including  
the ergoregion of rotating ABG black hole and also show how the charge parameter affects the ergoregion. It turns 
out that area of ergoregion depends both on the value of charge $Q$ and parameter $a$, and the area of ergoregion 
increase with either of the two parameters. 

We have also studied the collision of two different rest masses particles in the equatorial plane of rotating ABG 
black holes and also obtained an expression for the $E_{CM}$ for the particles. We demonstrate that the $E_{CM}$  
depends not only on the rotation parameter $a$ but also on the charge $Q$ of the black holes. This work is the 
generalization of a previous work \cite{Ghosh:2014mea}, where the analysis was restricted to the particles of 
the same rest mass particles moving in equatorial plane. Applying this general expressions, we realize that an 
infinite amount of energy can be obtained when the collision of the particles takes place in the vicinity of an 
event horizon of an extremal black hole and also on the inner horizon for the nonextremal black hole. Furthermore, 
we analyze the dependence of $E_{CM}$ on charge $Q$, and explicitly show the effect of $Q$ on the $E_{CM}$. Thus, 
we have estimated the $E_{CM}$ of two unequal mass particles for both the extremal and the nonextremal ABG black 
holes when one particle has the critical angular momentum. Furthermore, we analyze the dependence of $E_{CM}$ on 
charge $Q$.

\begin{acknowledgements}
S.G.G. would like to thank SERB-DST for Research Project Grant NO SB/S2/HEP-008/2014 and DST INDO-SA bilateral 
project DST/INT/South Africa/P-06/2016. M.A. thanks the University of KwaZulu-Natal and the National Research 
Foundation for financial support.
\end{acknowledgements}

\end{document}